\documentclass[preprint,5p]{elsarticle}

\usepackage{natbib}
\setcitestyle{authoryear,open={(},close={)}} 
\setcitestyle{citesep={;}} 
\usepackage{graphicx}
\usepackage{footnote}
\usepackage{epsfig}

\makesavenoteenv{tabular}
\makesavenoteenv{table}

\usepackage{url}
\usepackage{amssymb}

\usepackage{dcolumn}
\usepackage{bm}
\usepackage[switch]{lineno}

\usepackage{caption}
\usepackage{subcaption}

\journal{Astronomy and Computing}

\begin{document}\sloppy

\begin{frontmatter}


\title{\textit{Pipeline Collector}: gathering performance data for distributed astronomical pipelines}
\author{Alexandar P. Mechev$^a$ }
\ead{apmechev@strw.leidenuniv.nl}

\author{Aske Plaat$^b$}%
\author{J.B. Raymond Oonk$^a$$^,$$^c$}%
\author{Huib T. Intema$^a$}%
\author{Huub J.A. R\"ottgering$^a$}%

\date{\today}
\address{$^a$Leiden Observatory, Niels Bohrweg 2, 2333 CA Leiden, the Netherlands}
\address{$^b$Leiden Institute of Advanced Computer Science, Niels Bohrweg 1, 2333 CA Leiden, the Netherlands}
\address{$^c$ASTRON,  Oude Hoogeveensedijk 4, 7991 PD Dwingeloo, the Netherlands}

\begin{abstract}
Modern astronomical data processing requires complex software pipelines to process ever growing datasets. For radio astronomy, these pipelines have become so large that they need to be distributed across a computational cluster. This makes it difficult to monitor the performance of each pipeline step. To gain insight into the performance of each  step, a performance monitoring utility needs to be integrated with the pipeline execution. In this work we have developed such a utility and integrated it with the calibration pipeline of the Low Frequency Array, LOFAR, a leading radio telescope. We tested the tool by running the pipeline on several different compute platforms and collected the performance data. Based on this data, we make well informed recommendations on future hardware and software upgrades. The aim of these upgrades is to accelerate the slowest processing steps for this LOFAR pipeline. The \textit{pipeline\_collector} suite is open source and will be incorporated in future LOFAR pipelines to create a performance database for all LOFAR processing. 

\end{abstract}
\end{frontmatter}


%

\section{\label{sec:intro}Introduction }
Astronomical data often requires significant processing before it is considered ready for scientific analysis. This processing is done increasingly by complex and autonomous software pipelines, often consisting of numerous processing steps, which are run without user interaction. It is necessary to collect performance statistics for each pipeline step. Doing so will enable scientists to discover and address software and hardware inefficiencies and produce scientific data at a higher rate. To identify these inefficiencies, we have extended the performance monitoring package \textit{tcollector}\footnote{https://github.com/OpenTSDB/tcollector}\citep{tcollector}. The resulting suite, \textit{pipeline\_collector}, makes it possible to use \textit{tcollector} to record data for complex pipelines. We have used a leading radio telescope as the test case for the \textit{pipeline\_collector} suite. The discoveries made with our software will help remove bottlenecks and suggest hardware requirements for current and future processing clusters. We summarize our findings in Table \ref{table:results} in Section \ref{sec:results}.

Over the past two decades, processing data in radio astronomy has increasingly moved from personal machines to large compute clusters. Over this time, radio telescopes have undergone upgrades in the form of wide band receivers and upgraded correlators \citep{lofarcobalt,gmrt_upgrade}. In addition, several aperture synthesis arrays such as the Low Frequency Array \citep[LOFAR,][]{LOFAR}, Murchison Widefield Array \citep[MWA][]{MWA,mwa2} and MeerKAT \citep{meerkat} have begun observing the radio sky, leading to an increase of data rates by up to 3 orders of magnitude \citep{mwa_data_size,meerkat_size}.

As the data acquisition rate has increased, data size has entered the Petabyte regime, and processing requirements increased to millions of CPU-hours. In order for processing to match the acquisition rate, the data is increasingly processed at large clusters with high-bandwidth connections to the data. An important case where data processing is done at a high throughput cluster is the LOFAR radio telescope.

The LOFAR telescope is a European low frequency aperture synthesis radio telescope centered in the Netherlands with stations stretching across Europe. This aperture synthesis telescope requires significant data processing before producing scientific images \citep{lofar_prefactor,Wendy_bootes,tassesmirnov,oonk_2014}. In this work, we will use our performance monitoring utility, \textit{pipeline\_collector}\footnote{https://gitlab.com/apmechev/pipeline\_collector.git}, to study the first half of the LOFAR processing, the Direction Independent (hereafter  DI) pipeline. 

One major project for the LOFAR telescope is the Surveys Key Science Project (SKSP) \citep{lotss}. This project consists of more than 3000 observations of 8 hours each, 600 of which have been observed. These observations need to be processed by a DI pipeline, the results of which are calibrated by a Direction Dependent (DD) pipeline. The DI pipeline is implemented in the software package \textit{prefactor}\footnote{available at \protect\url{https://github.com/lofar-astron/prefactor}}. The \textit{prefactor} pipeline is itself split into four stages and implemented at the SURFsara Grid location at the Amsterdam e-Science centre \citep{SurfSara,mechev}. The automation and simple parallelization has decreased the run time per dataset from several days to six hours, making it comparable to the observation rate. To better understand and optimize the performance of the \textit{prefactor} pipeline, we require detailed performance information for all steps of the processing software. We have developed a utility to gather this information for data processing pipelines running on distributed compute systems. 

In this work, we will use the \textit{pipeline\_collector} utility to study the LOFAR \textit{prefactor} pipeline and suggest optimization based on our results. To test the software on a diverse set of hardware, we will set up the monitoring package on four different computers and collect data on the pipeline's performance. Using this data, we discuss several aspects of the LOFAR software which we present in Table \ref{table:results}. Finally we discuss the broader context of these optimizations in relation to the LOFAR SKSP project and touch on the integration of \textit{pipeline\_collector} with the second half of the data processing pipeline, the DD calibration and imaging. 
    
 \begin{table*}[!htb]
 \begin{center}
  \begin{tabular}{ l | p{130mm} }
    \hline
    Result \# & Description  \\ \hline
    \hline
    \textbf{R1} & Native compilation of the software performs comparably to pre-compiled binaries on two test machines. \\ \hline
    \textbf{R2} &  The processing steps do not appear to accelerate significantly on a faster processor or with larger cache size. \\ \hline
    \textbf{R3} & Both calibration steps (\textit{calib\_cal} and \textit{gsmcal\_solve}) show linear correlation between speedup and memory bandwidth. \\ \hline
    \textbf{R4} &  Disk read/write speed does not affect the completion time of the slowest steps. \\ \hline
    \textbf{R5} & Both calibration steps do not use large amounts of RAM despite processing data on the order of Gigabytes. \\ \hline
    \textbf{R6} & The \textit{calib\_cal} step can suffer up to 20\% of Level 1 Instruction Cache misses, while gsmcal only has 5\% of these misses. \\ \hline
    \textbf{R7} & Both calibration steps are impacted by Level 2 Cache eviction at comparable rates. \\ \hline
    \textbf{R8} & The \textit{calib\_cal} step stalls on resources 70\% of cycles while the gsmcal step only 30\% of them. \\ \hline
    \textbf{R9} & The \textit{calib\_cal} uses the CPU at full efficiency for only 10 \% of the CPU cycles. \\ \hline
    \hline
  \end{tabular}   
  \caption{A table of all the results presented in Section \ref{sec:results}.}
  \label{table:results}
   \end{center}
 \end{table*}

\subsection{Related Work}\label{sec:related}

Scientific fields that need to process large data sets employ some type of data processing pipelines.  Such pipelines include e.g. solar imaging \citep{solar_pipeline}, neuroscience imaging \citep{optimize_pipeline} and infrared astronomy \citep{herschel}. While these pipelines often log the start and finishing times of each step (using tools such as pegasus-kickstart \citep{kickstart}), they do not collect detailed time series performance data throughout the run. 

At a typical compute cluster the performance of every node in a distributed systems is monitored using utilities, such as Ganglia \citep{ganglia}. These tools only monitor the global system performance. 
If one is interested in specific processes, then the Linux procfs \citep{procfs} is used. The procfs system can be used to analyse the performance of individual pipeline steps. Likewise, the Performance API \citep[PAPI,][]{papi} is a tool which collects detailed low level information on the CPU usage per process. Collecting detailed statistics at the process level is required to understand and optimize the performance of the LOFAR pipeline and we will integrate PAPI into \textit{pipeline\_collector} in the future. Finally, DTrace\citep{dtrace} is a Sun Microsystems tool which makes it possible to write profiling scripts that access data from the kernel and can be used to monitor process or system performance at run time with minimal overhead. As DTrace was not installed on either of the processing clusters, we have not used it to monitor the pipeline's performance. 

The Linux procfs system and PAPI record data which is already made available by the Linux kernel. This option incurs insignificant overhead as it uses data the kernel and processor already log. Likewise PAPI reads performance counters that the CPU automatically increments during processing. These profiling utilities can run concurrently with the scientific payload without using more than 1-2\% of system resources. Their low overhead is why we choose to use them to collect performance data. 

Other tools for performance analysis such as Valgrind \citep{valgrind} collect very detailed performance information. This comes at the expense of execution time: running with Valgrind, the processing time slows by up to two orders of magnitude. As such, we do not use Valgrind along the LOFAR software. 

\section{Measuring LOFAR Pipeline performance with pipeline\_collector}\label{sec:methods}
We developed the package \textit{pipeline\_collector} as an extension of  the performance collection package \textit{tcollector}. \textit{pipeline\_collector} makes it possible to collect performance data for complex multi-step pipelines. Additionally, it makes it easy to record performance data from other utilities. A performance monitoring utility that we plan to integrate in the future are the PAPI tools described in section \ref{sec:related}. The resulting performance data was recorded in a database and analyzed. For our tests, we used the LOFAR \textit{prefactor} pipeline, however with minor modifications, any multi-step pipeline can be profiled. 

\textit{tcollector} is a software package that automatically launches 'collector' scripts. These scripts are  sample the specific system resource and send the data to the main tcollector process. This process then sends the data to the dedicated time series database. We created custom scripts to monitor processes launched by the \textit{prefactor} pipeline (\ref{sec:customcollectors}). 

In this work, we use a sample LOFAR SKSP data set as a test case. A particular focus was to understand the effect of hardware on the bottlenecks of the LOFAR data reduction. To gain insight into the effect of hardware on \textit{prefactor} performance, the data was processed on four different hardware configurations (Table \ref{tab:nodes}). As typical upgrade cycle for cluster hardware is five years, our results will be used to select optimal hardware for future clusters tasked with LOFAR processing.  

\subsection{\textit{Prefactor} Pipeline}

The LOFAR \textit{prefactor} pipeline \citep{lofar_prefactor} is a software pipeline that performs direction independent calibration using the LOFAR software. The LOFAR software stack is a software package containing commonly used processing software used by LOFAR pipelines \citep{cookbook,lofar_NDPPP}. These tools are built and maintained by ASTRON\footnote{ASTRON: Netherlands Institute for Radio Astronomy, url{https://www.astron.nl/}}. 

\begin{figure}
    \includegraphics[width=0.95\linewidth]{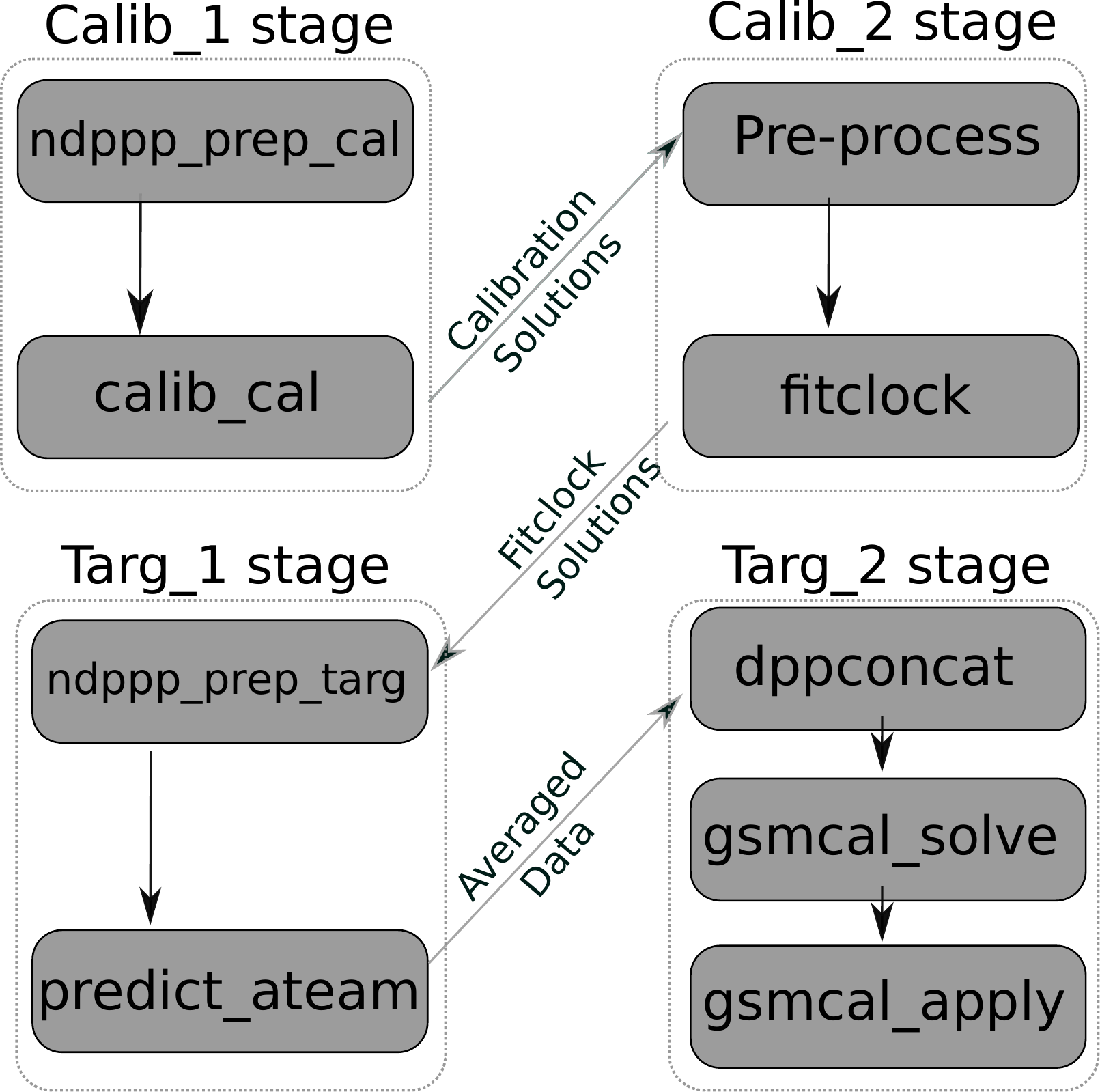}
      \caption{The four processing stages that make up the prefactor pipeline. The Calibrator stages (top) process a known bright calibrator to obtain the gain for the LOFAR antennas. The Target stages (bottom) process the scientific observation to remove Direction Independent effects. The \textit{pref\_cal1} and \textit{pref\_targ1} stages are massively parallelized across nodes without the need of an interconnect. The \textit{pref\_cal2} step runs only on one node, while \textit{pref\_targ2} is parallelized on 25 nodes. As the LOFAR software does not use MPI, we can run each processing job in isolation. }
	\label{fig:four_steps_box}
\end{figure}

The \textit{prefactor} pipeline performs a sequence of four stages, namely the calibrator and target calibration. The first half of \textit{prefactor} processes data from a calibration source and the second half processes a science target. Altogether, this processing takes six hours on a high-throughput cluster. The final result is a data-set ready for creating  images of the sky at radio wavelengths. Figure \ref{fig:four_steps_box} shows a graphical view of the prefactor pipeline's Calibrator and Target stages.

The Calibrator stage consists of the \textit{ndppp\_prep\_cal} and the \textit{calib\_cal} step. The former flags radio interference and averages the data, and the latter performs gain calibration on a bright calibration source. It is followed by the \textit{fitclock} step which fits a clock-TEC model to the calibration solutions \citep{lofar_prefactor}.  

\begin{figure} 
    \includegraphics[width=0.95\linewidth]{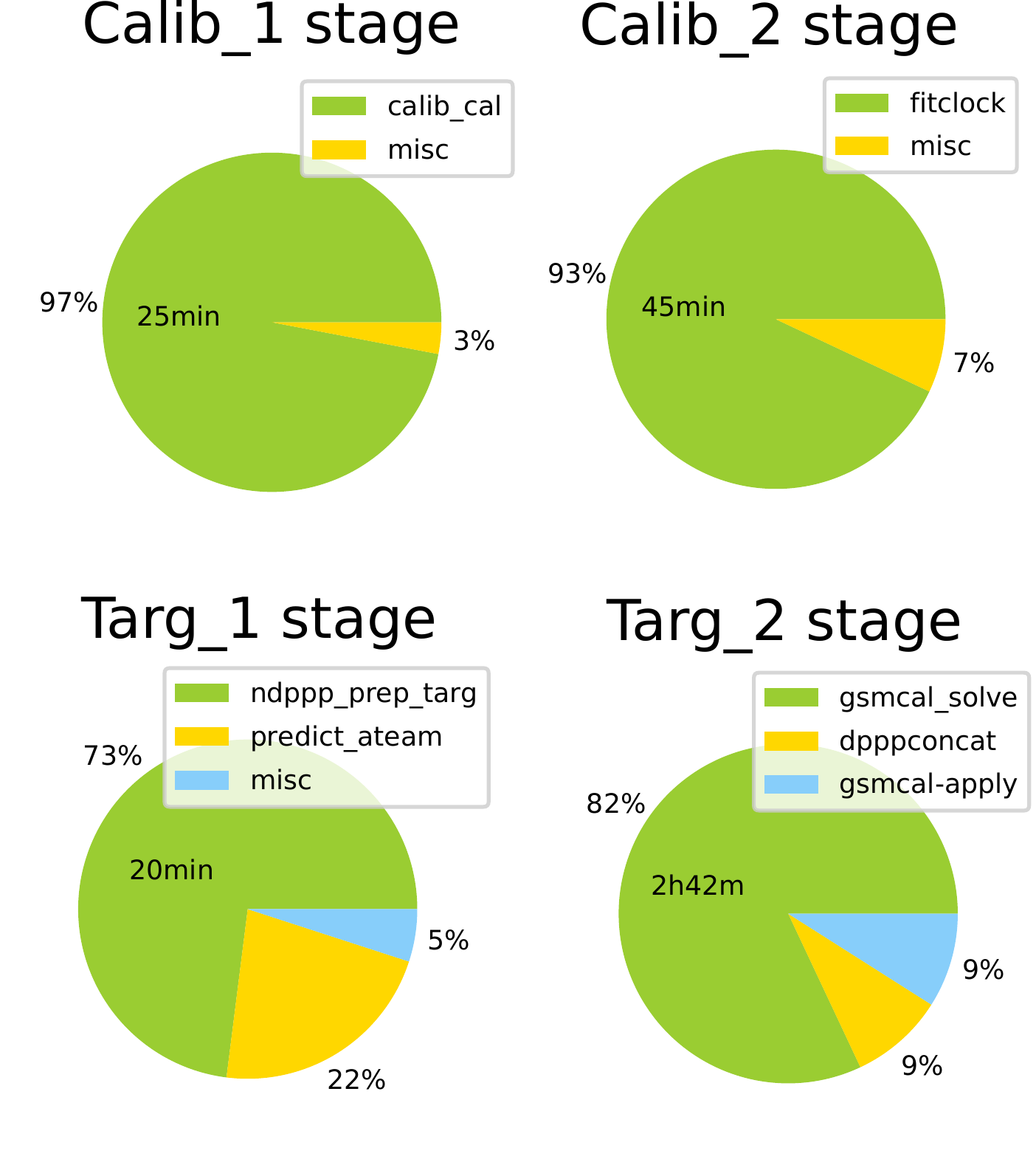}
      \caption{Portion of processing time taken by each step for the four prefactor stages, as reported by the Prefactor software. For each stage, the majority of processing time is spent during one or two steps. This is due to the fact that each prefactor stage also has intermediate book-keeping steps explicitly included in the pipeline. For each pipeline stage, the mean processing time for the longest running steps at SURFsara is also indicated. It should be noted that while faster, pref\_cal1 runs 10 times as many jobs as pref\_targ2. }
	\label{fig:4_steps_pies}
\end{figure}

The Target stage consists of a \textit{ndppp\_prep\_targ} step, \textit{predict\_ateam}, \textit{gsmcal\_solve} and \textit{gsmcal\_apply} steps. The first two of these steps flag and average the target data and calculate contamination by bright off-axis radio sources. The \textit{gsmcal\_solve} step determines phase solutions for each antenna using a model of the target and the results of the \textit{ndppp\_prep\_targ} step. Finally, the \textit{gsmcal\_apply} step applies these solutions to the target data. Figure \ref{fig:4_steps_pies} shows the percentage of time spent by these steps for the four \textit{prefactor} stages.

\subsection{Performance suite}
Cluster performance is frequently monitored using utilities such as Ganglia \citep[][discussed in Section \ref{sec:related}]{ganglia}. These tools cannot access individual processes and thus cannot collect data on a per-process basis. To collect such data, each process launched by the active pipeline step needs to be profiled individually. Our utility is designed to gather such performance data.

Our monitoring package, \textit{pipeline\_collector} adds custom performance collectors (\ref{sec:customcollectors}) to the performance collection framework \textit{tcollector}. We use these collectors to monitor individual pipeline steps as defined by the user\footnote{https://gitlab.com/apmechev/pipeline\_collector.git}. The tools attach to processes launched by the pipeline and record performance data at a one second interval. This sampling frequency is at high enough resolution to detect trends and anomalies in hardware utilization, and still result in a reasonable database size. The performance data is uploaded to a remote time series database, OpenTSDB \citep{opentsdbsite}. Details on the data collection can be found in  \ref{appendix1}.

\subsubsection{Performance API}\label{sec:papi2}

The time-series database is also used to collect low-level CPU information for each process. This information is collected by the PAPI interface (discussed in Section \ref{sec:related}). This was done through the papiex utility\footnote
{Available at https://bitbucket.org/minimalmetrics/papiex-oss} \citep{papiex}. This utility records the CPU's internal performance counters. A CPU's performance counters record information on how efficiently the software uses the CPU's resources. The results from this test are detailed in Section \ref{sec:PAPI}.

\subsection{Test Hardware}

In order to study the effect of different hardware configurations on the performance of LOFAR processing, the \textit{prefactor} pipeline was run on four different sets of hardware. The four machines tasked with processing LOFAR data comprised nodes at three computational clusters and a personal computer. The tests were run while the systems were idle to make sure there is no interference of other software with the LOFAR processing. Table \ref{tab:nodes} details the specifics of the four test machines.

 \begin{table*}
 \begin{center} 

  \begin{tabular}{ l | c | c | c | c | c | c }
    \hline
    Location & CPU Speed (MHz) & CPU Model & Micro-architecture & Cache Size & RAM Speed\footnotemark & Disk Speed\footnotemark \\ \hline
    \hline
    Leiden & 2200 & E5-4620 & Sandy Bridge & 16 MB  & 1.4 GB/s & 99.7 MB/s \\ \hline
    SURFsara & 2500 & E5-2680 & Sandy Bridge & 30 MB  & 2.5 GB/s & 65.4 MB/s \\ \hline
    Hatfield &  2900 & E5-2660 & Sandy Bridge  & 20 MB   & 2.4 GB/s & 155 MB/s\\ \hline
    Laptop & 3300 & E3-1505M & Skylake &  8 MB  & 4.7 GB/s & 822 MB/s\\ \hline
    \hline
  \end{tabular}   
  \caption{CPU, Cache, RAM and Storage specifications of the four test machines. The tested machines span a factor of 1.5x in CPU speed, 4x in cache and RAM Speed and $~$10x in Disk speed.}
  \label{tab:nodes}
  \end{center}
 \end{table*}

\section{LOFAR Prefactor Test Case}\label{sec:results}
  
With the test set described in Section \ref{sec:methods}, we aim to understand processing bottlenecks in the \textit{prefactor} pipeline and make informed decisions on future hardware and software upgrades. To do so, we processed a sample observation at institutes that typically process LOFAR data. 

From the data collected by processing the sample observation, we determined the slowest pipeline steps. These steps were the \textit{calib\_cal} and \textit{gsmcal\_solve}, seen in Figure \ref{fig:4_steps_pies}. The \textit{calib\_cal} step is implemented by the software \texttt{bbs-reducer} \citep{cookbook,bbs_selfcal} and the \textit{gsmcal\_solve} step is implemented by \texttt{NDPPP} \citep{cookbook,lofar_NDPPP}. Both \texttt{bbs-reducer} and \texttt{NDPPP} are part of the LOFAR software suite.

\begin{figure*}
  \centering
   \begin{subfigure}{.45\linewidth}
    \includegraphics[width=\textwidth]{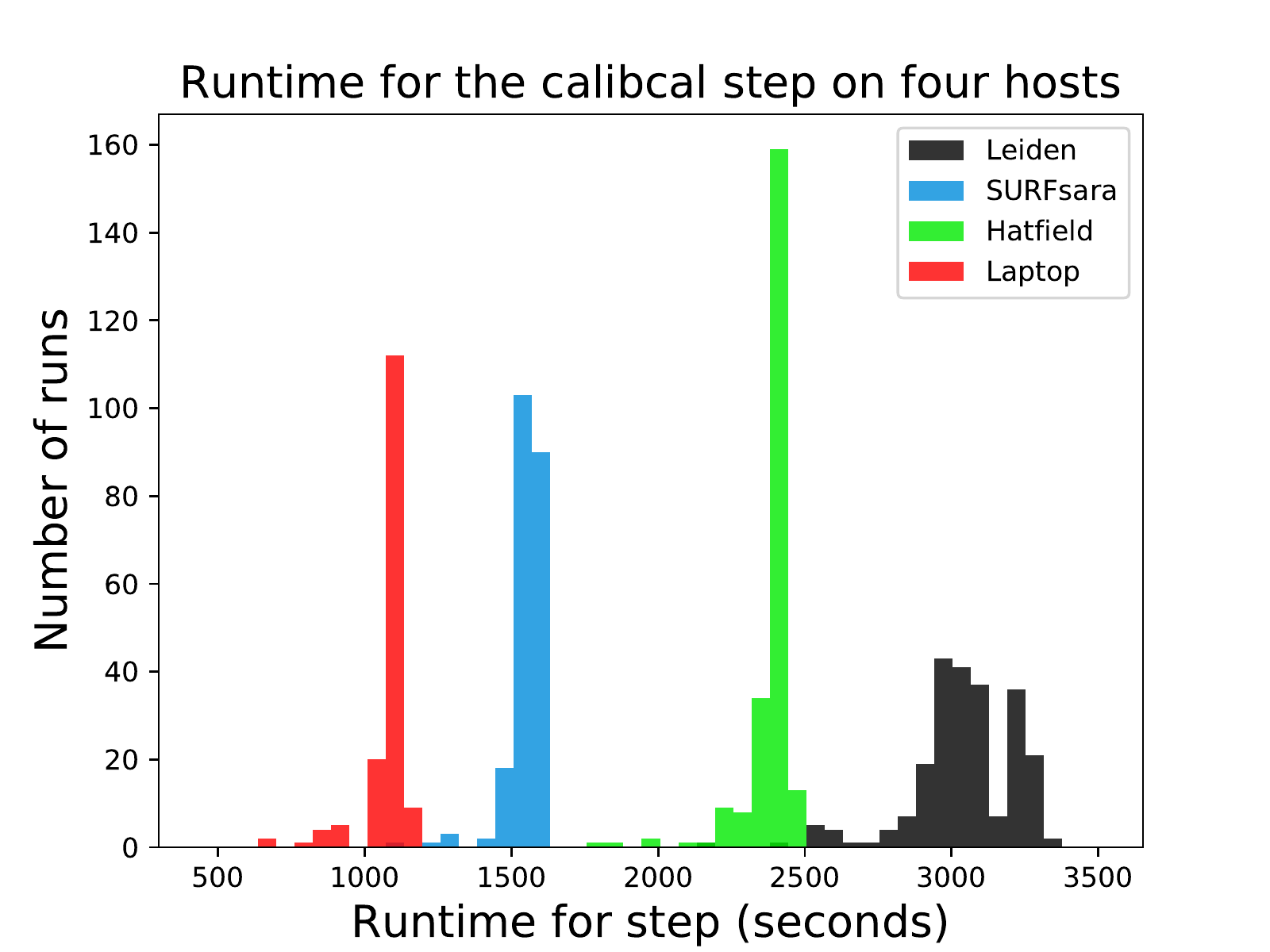}
    \caption{calib\_cal}
	\label{calib_cal_bar}
 \end{subfigure}%
 \begin{subfigure}{.45\linewidth}
  \includegraphics[width=\textwidth]{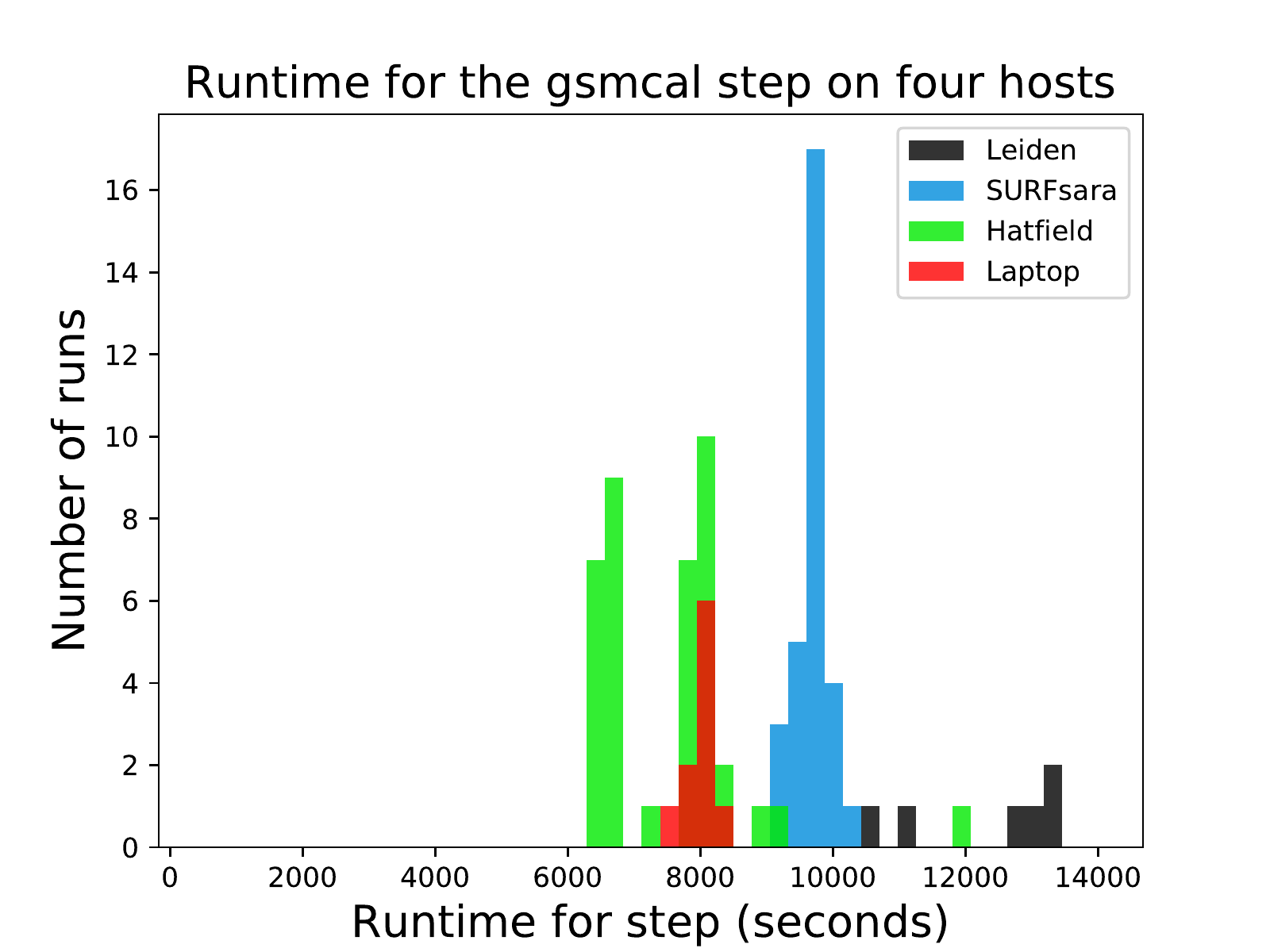}
  \caption{\textit{gsmcal\_solve} }
  \label{gsmcal_bar}
 \end{subfigure}
  \caption{Job completion times for \textit{calib\_cal} and \textit{gsmcal\_solve} steps tested on four hardware setups. The \textit{calib\_cal} step ran 244 times. The \textit{gsmcal\_solve} ran 24 times as the data is concatenated from 244x1 to 24x10 sets.  The step with the longest latency is \textit{gsmcal\_solve} while \textit{calib\_cal} consumes a comparable number of core-hours over 244 jobs. } 
  \label{calibcal_fitclock_bars}
\end{figure*}

We collected performance statistics using the \textit{pipeline\_collector} suite as discussed in Section \ref{sec:methods}. The runtime of the slowest \textit{prefactor} steps on the four machines is shown in figure \ref{calibcal_fitclock_bars}. The results discovered using \textit{pipeline\_collector} are listed in Table \ref{table:results} and discussed in Section \ref{sec:res_tcoll}. Using the PAPI interface (discussed in Section \ref{sec:papi2}) CPU performance data was collected. The results from this test are detailed in Section \ref{sec:PAPI}.

We will present a number of insights into the performance of the LOFAR software collected by the profiling suite. The results are presented in Table \ref{table:results} and are grouped in three main areas. The effect of compilation on the runtime was result \textbf{R1}. The set of results \textbf{R2}, \textbf{R3}, \textbf{R4} and \textbf{R5} were obtained using the \textit{pipeline\_collector} package. Results  \textbf{R6}, \textbf{R7}, \textbf{R8} and \textbf{R9} were collected with the PAPI package, which will be integrated into \textit{pipeline\_collector} in the future.  

\subsection{Pre-compiled vs native compilation}
 \footnotetext[7]{benchmarked using $dd$}
 \footnotetext{sequential disk read, benchmarked using \texttt{fio} - flexible I/O tester: \newline \newline \texttt{fio --randrepeat=1 --ioengine=libaio --direct=1 --gtod\_reduce=1 --name=test --filename=test --bs=4k --iodepth=256 --size=4G --readwrite=read --ramp\_time=4} \newline } 

The performance trade-off between pre-compiled and native compilation was studied first. The majority of the processing for the LOFAR SKSP Project \citep{lotss} is done at the SURFsara \textsc{gina} cluster in Amsterdam. This location is part of the European Grid Initiative (EGI)\citep{SurfSara}. At this location, software is deployed by compiling on a virtual machine and mounting it on all worker nodes through the CernVM FileSystem (CVMFS) service \citep{cvmfs}. The CVMFS server allows any client to mount a fully compiled LOFAR installation, making it easy to distribute and version control the software within and outside of SURFsara. An alternative is to locally compile the LOFAR packages on each cluster. The performance of the natively compiled\footnote{The software was compiled using \texttt{-march=native} and \texttt{-O3} compilation flags. On the laptop, gcc resolves \texttt{-march=native} as \texttt{broadwell}. The CVMFS installation resolves \texttt{-march=native} as \texttt{core-avx-i}.} vs CVMFS installations was compared on the laptop test machine using \textit{pipeline\_collector}. In order to validate this result, the two compilations of the same software were also tested at the Data Science Lab at the Leiden Institute of Advanced Computer Science (LIACS)\footnote{https://www.universiteitleiden.nl/en/science/computer-science/about-us/our-facilities}. 
 
\begin{figure*}
  \centering
     \begin{subfigure}{.45\linewidth}
    \includegraphics[width=\linewidth]{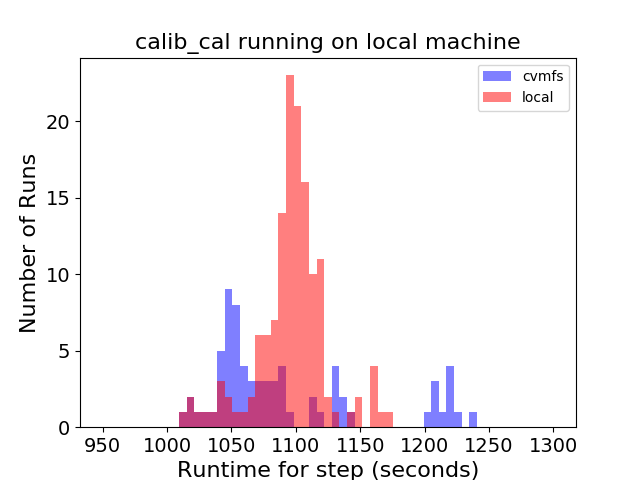}
    \caption{Two compilation options on a Laptop}
	\label{fig:cmvfs1_calib}
    \end{subfigure}%
    \begin{subfigure}{.45\linewidth}
     \includegraphics[width=\linewidth]{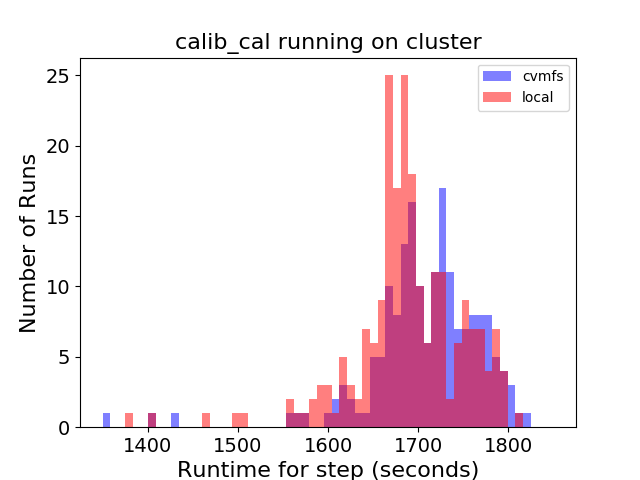}
    \caption{Two compilation options on cluster node}
	\label{fig:cmvfs2_calib}
    \end{subfigure}%
      \caption{Difference in processing time for \textit{calib\_cal} when compiled remotely and natively. \textit{calib\_cal} was run 244 times with the native software and 40 times with the CVMFS compilation. Two tests were done, one on the personal laptop (\ref{fig:cmvfs1_calib}) and one on a cluster node at the LIACS Data Science Lab (\ref{fig:cmvfs2_calib}). The test on a cluster node shows no significant difference in runtime between compilation options. The laptop test suggests that the remotely compiled software may run 5\% faster than the local compilation. } 
	\label{fig:cvmfs_native}
\end{figure*}

\begin{figure*}
  \centering
     \begin{subfigure}{.45\linewidth}
    \includegraphics[width=\linewidth]{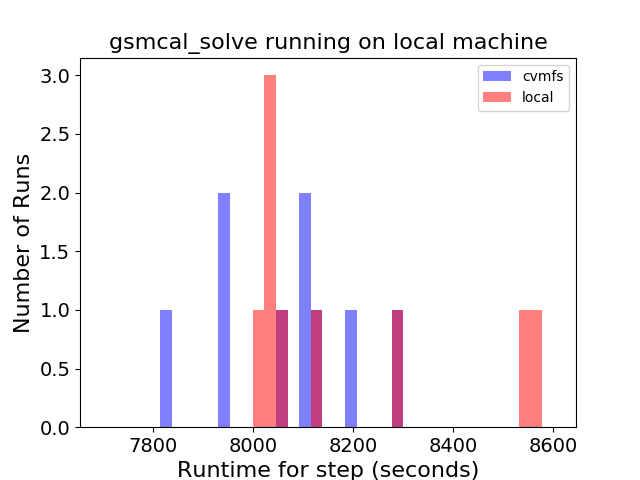}
    \caption{Two compilation options on a Laptop}
	\label{fig:cmvfs1_gsmcal}
    \end{subfigure}%
    \begin{subfigure}{.45\linewidth}
     \includegraphics[width=\linewidth]{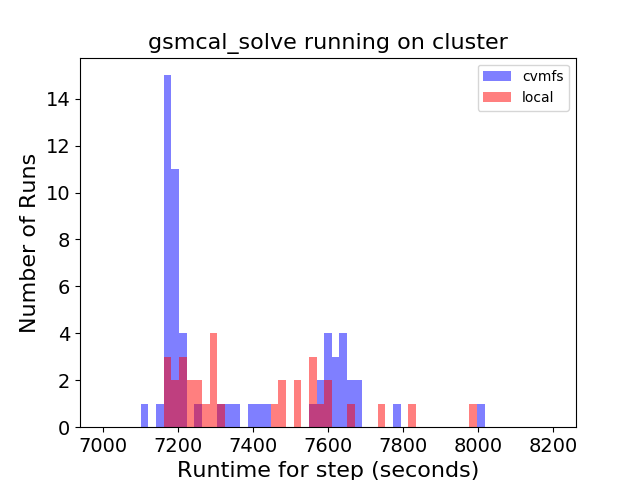}
    \caption{Two compilation options on cluster node}
	\label{fig:cmvfs2_gsmcal}
    \end{subfigure}%
      \caption{Difference in processing time for \textit{gsmcal\_solve} when compiled remotely and natively. \textit{gsmcal\_solve} was run 50 times with the native software and 120 times with the CVMFS compilation. Two tests were done, one on the personal laptop (\ref{fig:cmvfs1_gsmcal}) and one on a cluster node at the LIACS Data Science Lab (\ref{fig:cmvfs2_gsmcal}). Just like with the \textit{calib\_cal} step, the \textit{gsmcal\_solve} step also doesn't accelerate significantly when natively compiled.} 
	\label{fig:cvmfs_native_gsm}
\end{figure*}

An interesting discovery is that the LOFAR software did not process data faster when compiled natively. This is despite the fact that the local install was compiled with advanced processor instructions available on the host machine.  Figure \ref{fig:cvmfs_native} shows a histogram of its processing time with the two different compilation options for the \textit{calib\_cal} software running on the sample dataset. The same test was done for the software performing the gain calibration (\textit{gsmcal\_solve}), seen in Figure \ref{fig:cvmfs_native_gsm}. The result of this experiment is shown in Figures \ref{fig:cmvfs1_calib} and \ref{fig:cmvfs1_gsmcal}.  The software compiled at SURFsara showed a minor improvement for the \textit{calib\_cal} step on the laptop machine, however this improvement is not seen on the computational cluster node. 

Overall, the software for both steps show no significant improvement when compiled natively. This is result \textbf{R1} in Table \ref{table:results}. The second run at LIACS also confirms this result for both steps (Figures \ref{fig:cmvfs2_calib} and \ref{fig:cmvfs2_gsmcal}). This result suggests that the slowest \textit{prefactor} steps are not optimized for modern processors. 


\subsection{Prefactor Runtime and Hardware Parameters}\label{sec:res_tcoll}

Next, we studied the dependence of runtime on different hardware parameters. With software that collects per-step performance statistics for the LOFAR pipeline, the dependence of the pipeline processing on hardware performance can be easily profiled and studied. Using \textit{pipeline\_collector} we determined the pipeline's slowest steps with respect to different hardware parameters. 

\begin{figure}[ht!]
  \centering
    \includegraphics[width=0.7\linewidth]{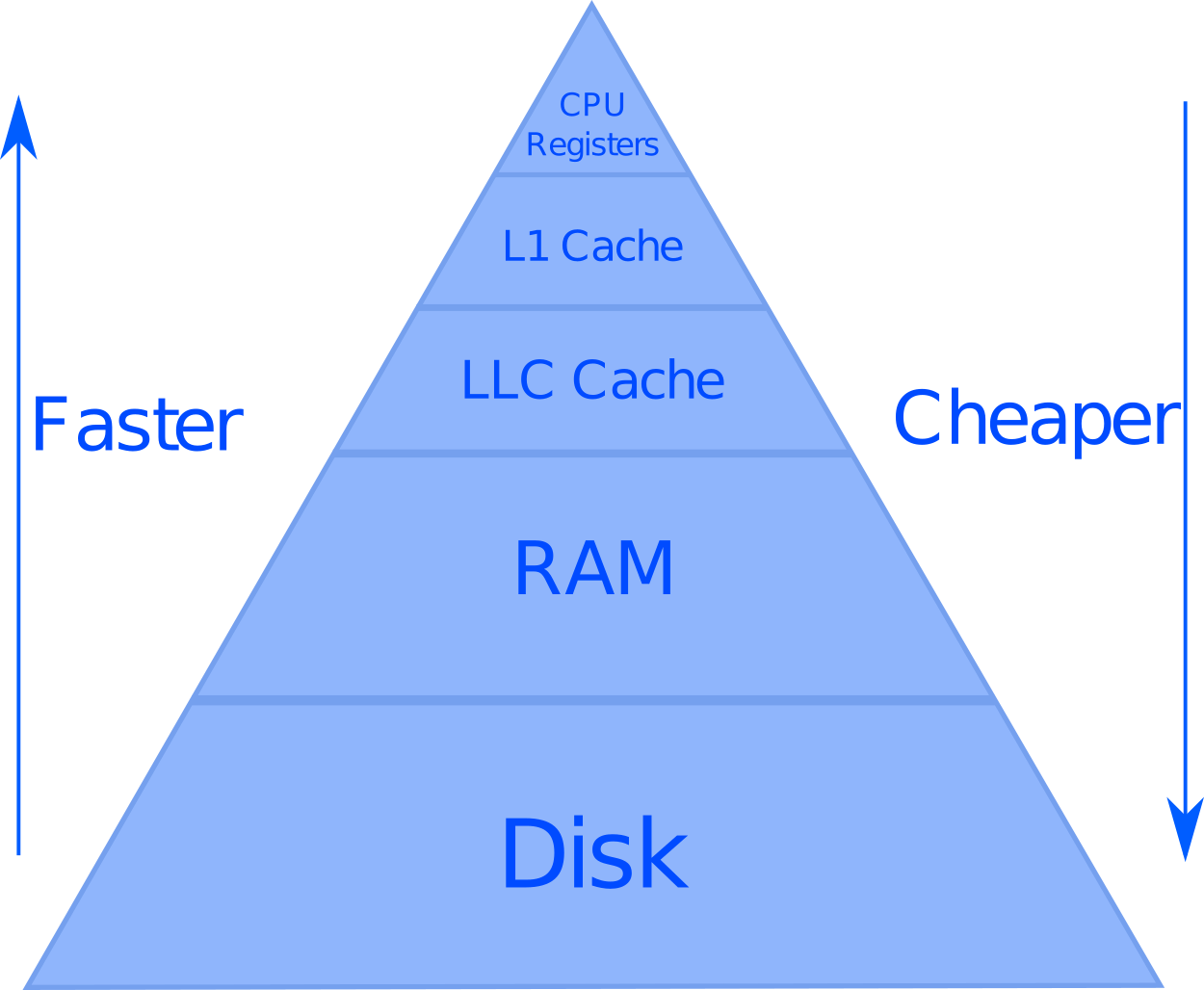}
      \caption{A model of the memory hierarchy, as described in \citep{goto2008anatomy}. }
	\label{fig:mem_hiearch}
\end{figure}

The system parameters studied here are the CPU speed, memory throughput, cache size and disk speed. Modern computers can have a complex memory hierarchy as demonstrated in Figure \ref{fig:mem_hiearch} \citep{mem_hiearch}. This is due to the cost trade-off between memory size and memory speed. Because of this trade-off, the full dataset is stored on disk, while the working set is placed in RAM. This is the data that the processor needs to access at the current time \citep{workingset}. The most frequently accessed parts of the data are stored in the CPU cache, which evicts the oldest data when full \citep{cache_eviction}. 

The CPU processing speed is faster than the RAM latency, so a hierarchy of caches exist. Caches store small subsets of the working set and have a fast connection to the processor. The fastest data link is between the CPU and the L1 Cache, with the link to RAM being slower and the disk read speed slower still. The limited memory capacity of the different levels of the memory hierarchy  as well as the throughput between them will lead to performance bottlenecks. These bottlenecks will lead to the processor waiting on memory. Such stalls lead to longer processing times.

\subsubsection{CPU}
The CPU speed is usually the primary factor determining how fast computations can be made. In general, a faster CPU will result in faster data processing. 

\begin{figure}
\begin{minipage}[b]{0.45\textwidth}
\begin{subfigure}[b]{\linewidth}
    \includegraphics[width=\textwidth]{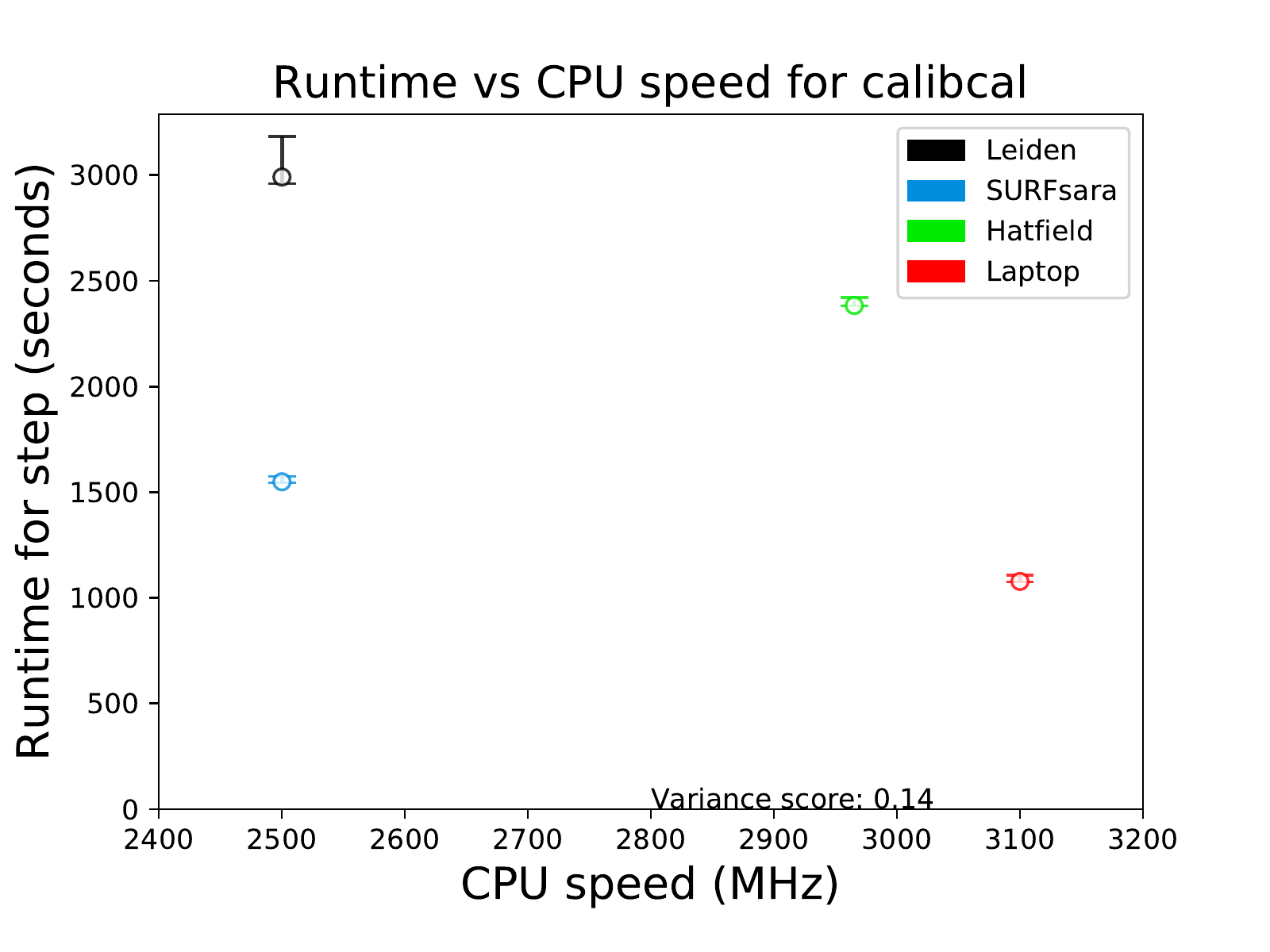}
      \caption{calib\_cal }
	\label{calib_cal_CPU}
 \end{subfigure}%

\vspace*{5mm} 
\begin{subfigure}[b]{\linewidth}
    \includegraphics[width=\textwidth]{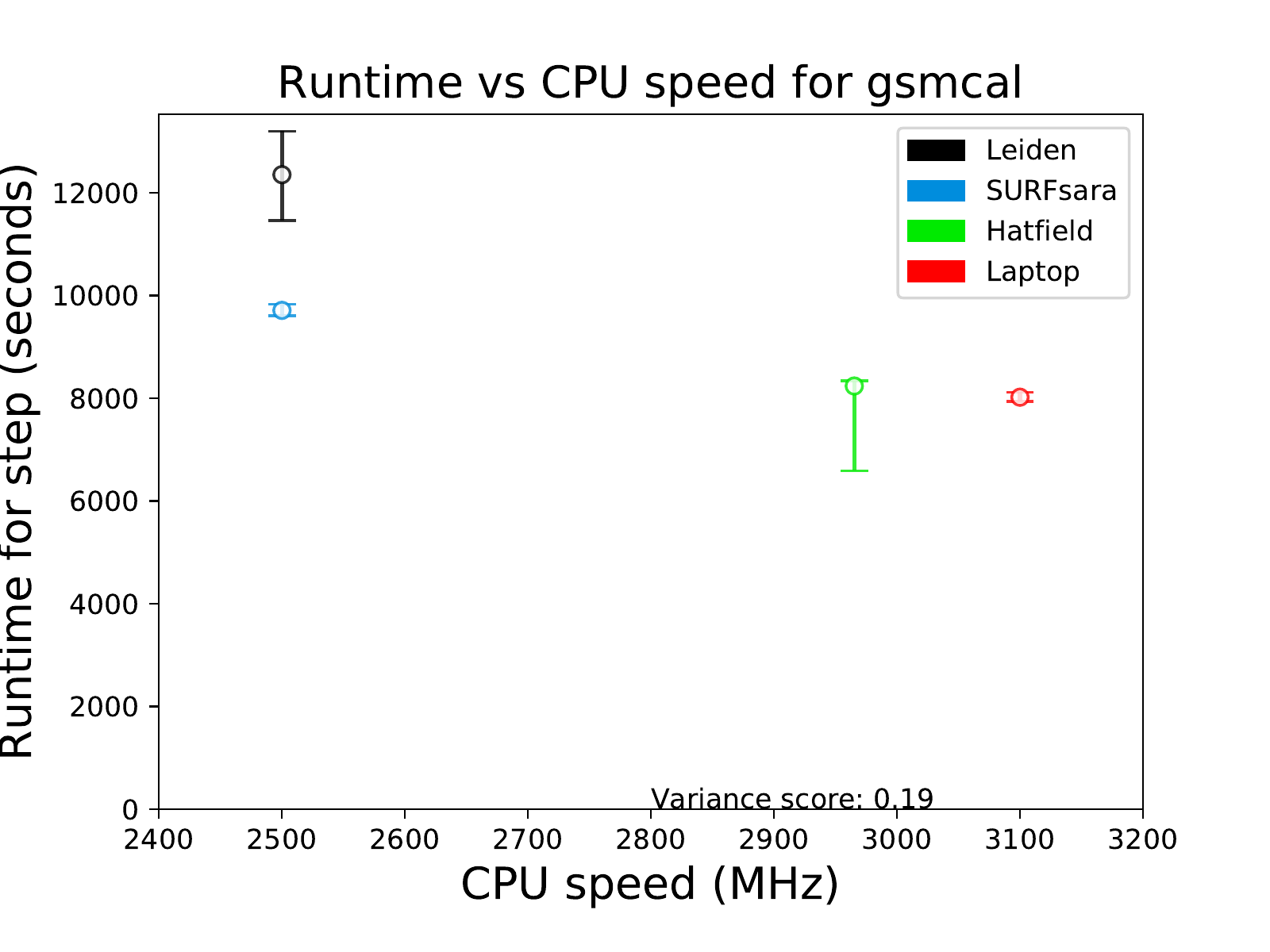}
      \caption{\textit{gsmcal\_solve}}
	\label{gsmcal_CPU}
 \end{subfigure}
 \label{CPU_3_steps}
 \caption{Performance of the bottleneck steps compared with the CPU speeds of the four test machines. The values are the mean of 244 runs (Standard prefactor run) and the error bars show the 1-sigma of the distribution of the run time. } 
\end{minipage}
\hfill        %
\end{figure}

However, Fig. \ref{calib_cal_CPU} shows that the runtime of the calibration of the calibrator does not strongly depend on the CPU frequency. While the test nodes at SURFsara and Leiden run at the same CPU frequency, running on a cluster node at SURFsara takes half the time as on a node at Leiden. Even more surprisingly, the \textit{gsmcal\_solve} step does not benefit significantly from a faster CPU, despite being the most computationally heavy \textit{prefactor} step (\textbf{R2}). This step does the gain calibration on the target field using the StEFCal algorithm \citep{stefcal}. Figure \ref{gsmcal_CPU} shows only a slight improvement over faster CPU clock speeds for both steps. The correlation between completion time and CPU speed is similar for both steps.

\subsubsection{Cache}
The CPU has a hierarchy of caches consisting of Level 1, Level 2 Cache and LLC Cache. For the four processors tested, the Level 1 and 2 caches were all the same size, thus the only difference is the Last Level Cache (LLC or just Cache in Figure \ref{fig:mem_hiearch}). This cache stores data needed by the CPU, so the larger it is, the less the processor needs to wait for RAM to return data. 

\begin{figure}
\begin{minipage}[b]{0.45\textwidth}
\begin{subfigure}[b]{\linewidth}
    \includegraphics[width=\textwidth]{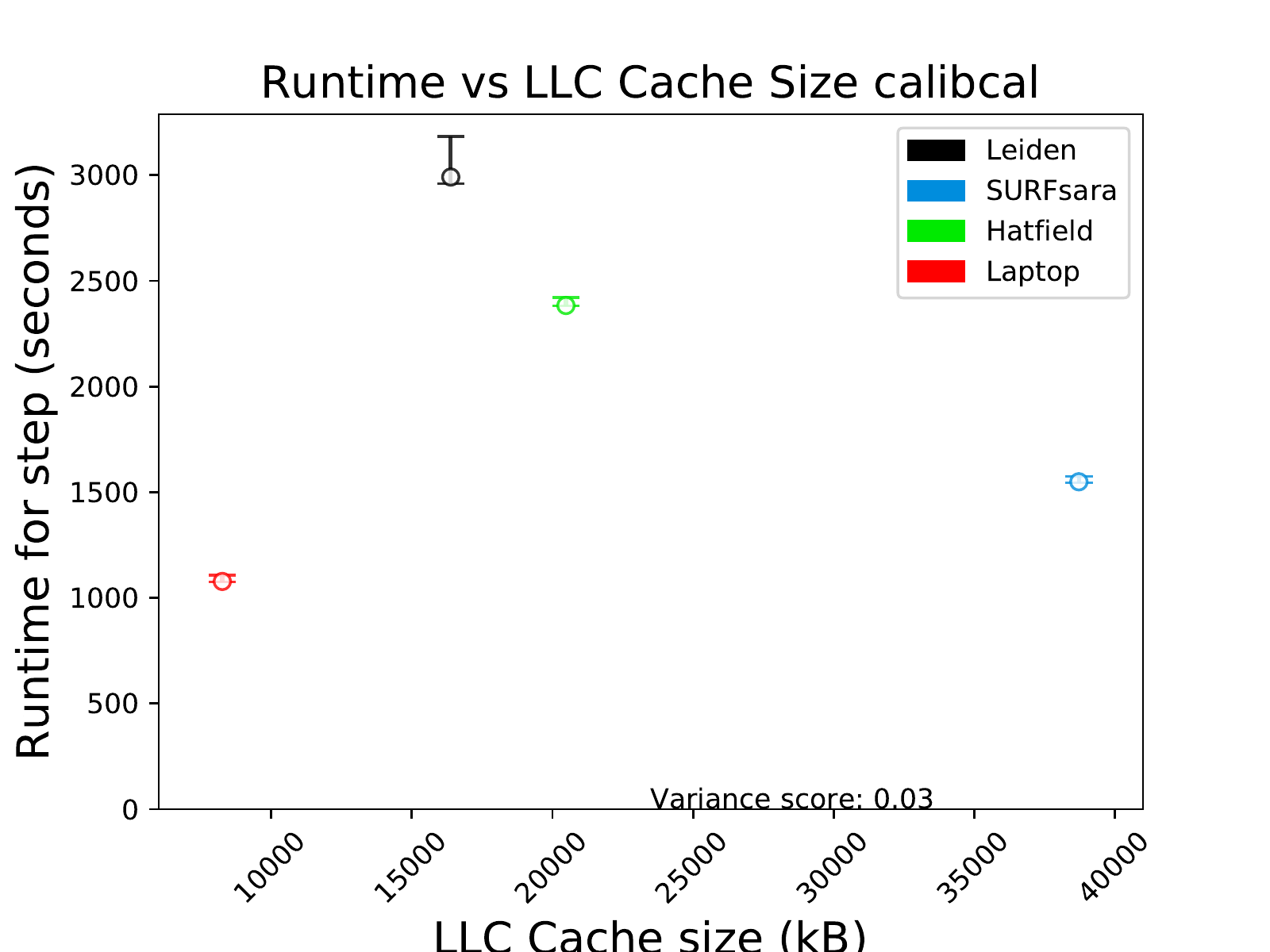}
      \caption{calib\_cal }
	\label{calib_cal_cache}
 \end{subfigure}%

\vspace*{5mm} 
\begin{subfigure}[b]{\linewidth}
    \includegraphics[width=\textwidth]{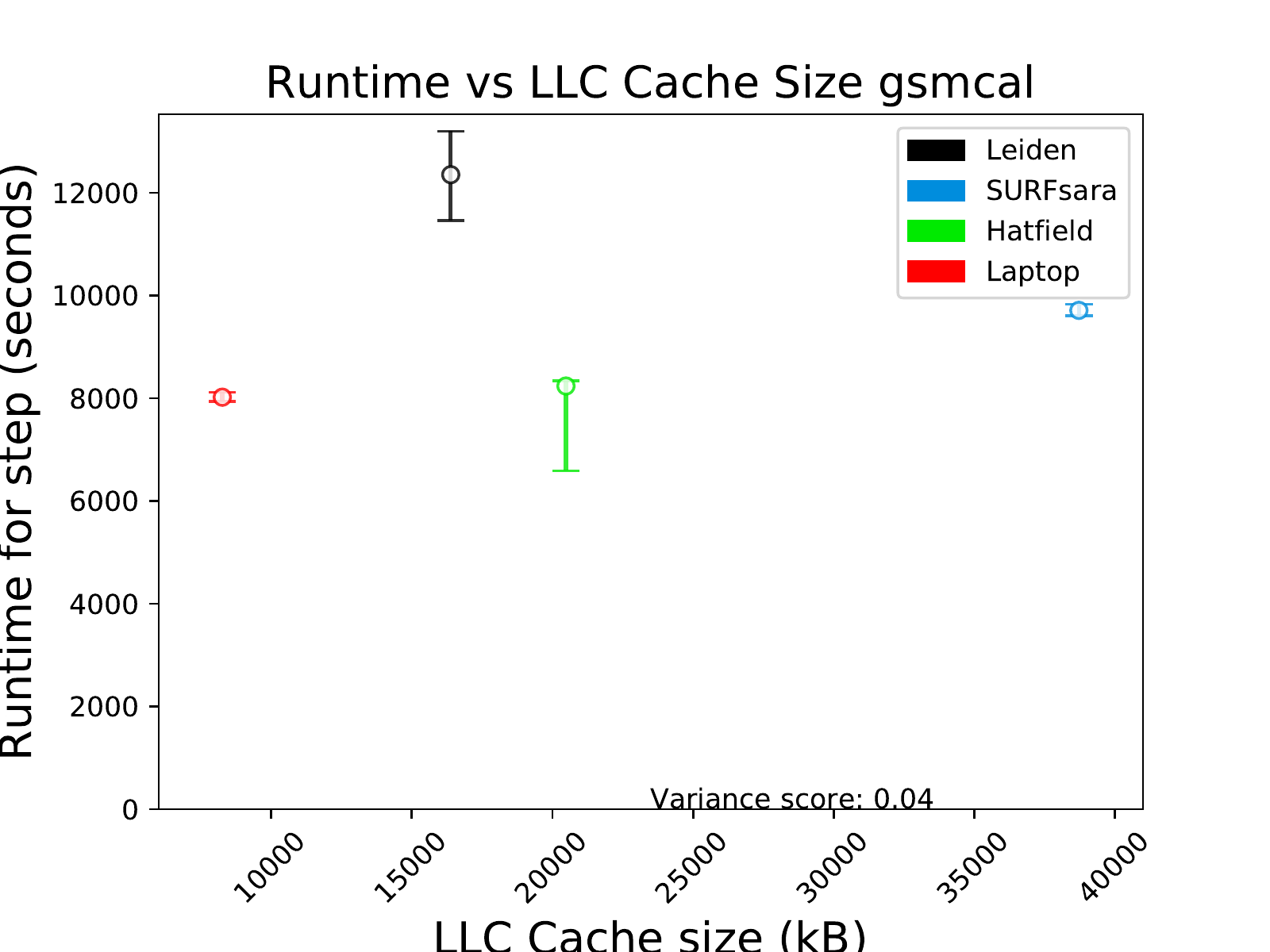}
      \caption{\textit{gsmcal\_solve}}
	\label{gsmcal_cache}
 \end{subfigure}
 \label{cache_3_steps}
 \caption{Performance of the two bottleneck steps with respect to Last Level Cache size. The \textit{gsmcal\_solve} step shows no trend between cache size and completion time. The \textit{calib\_cal} step runs the fastest on the machine with the smallest cache.  } 
\end{minipage}
\end{figure}

In general, numerical codes benefit from larger cache sizes \citep{skadron1999branch,goto2008anatomy}. Interestingly,  figure \ref{gsmcal_cache} suggests that the \textit{gsmcal\_solve} step does not exclusively depend on larger cache \textbf{R3} (Table \ref{table:results}). On the machines with a larger cache, the \textit{gsmcal\_solve} step completed processing as quickly as on the machines with smaller cache, even down to 8MB.  

\subsubsection{RAM Bandwidth}

If the entire data set does not fit into cache, the software needs to transfer data from RAM to the CPU. In these cases,  \textit{prefactor} benefits from a fast bandwidth between the cache and RAM. For this study, the RAM throughput was benchmarked\footnote{Using the command \texttt{\$> dd if=/dev/zero of=/dev/shm/test  bs=1M count=2048}}. This command copies dummy data into system memory. As this utility exists on all Unix systems, this is a standardized benchmark of the RAM performance. 

\begin{figure}
\begin{minipage}[b]{0.45\textwidth}
\begin{subfigure}[b]{\linewidth}
    \includegraphics[width=\textwidth]{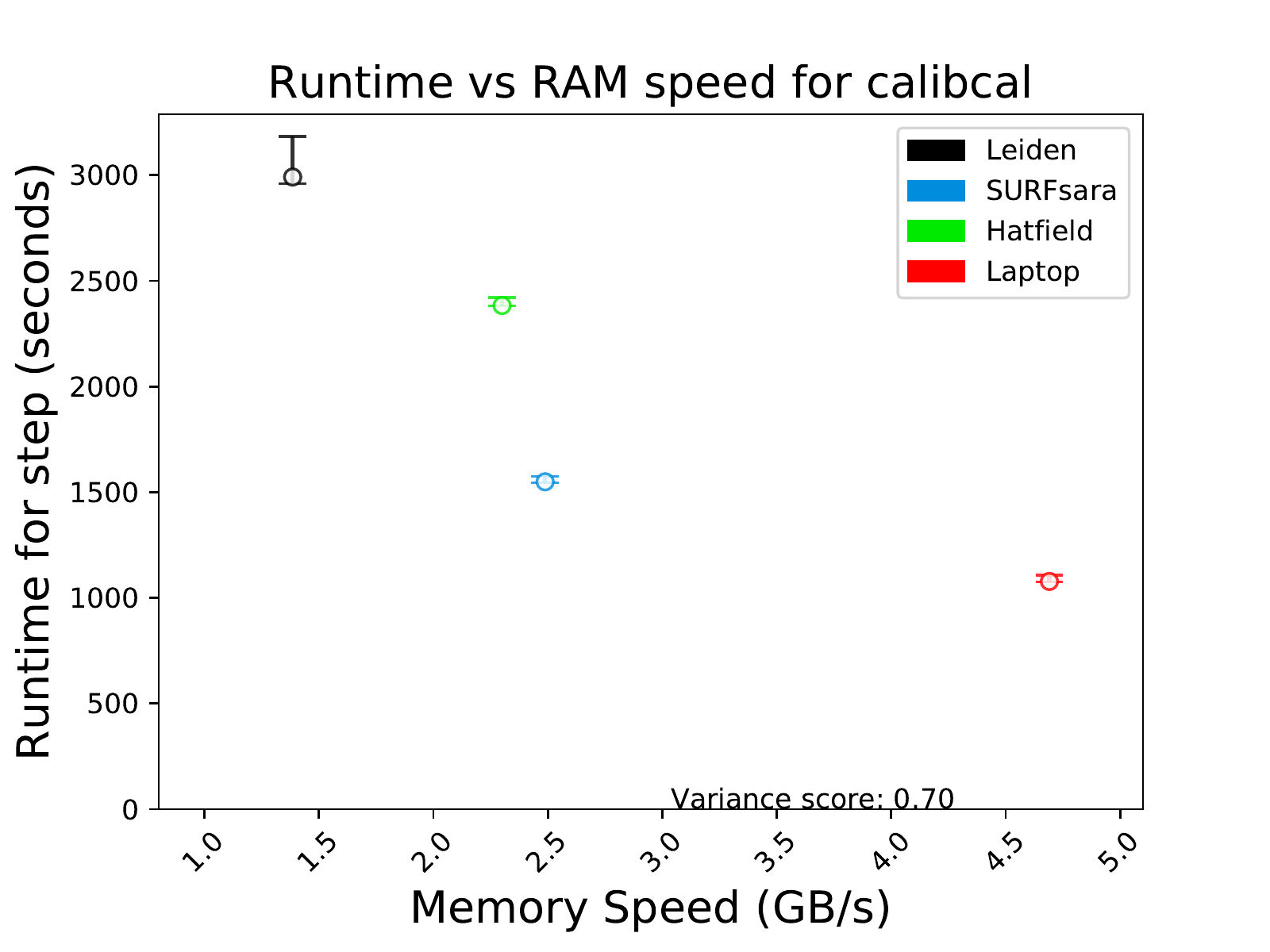}
      \caption{\textit{calib\_cal}}
	\label{calib_cal_RAM}
 \end{subfigure}%
 \vspace*{5mm} 

  \begin{subfigure}{\linewidth}
    \includegraphics[width=\textwidth]{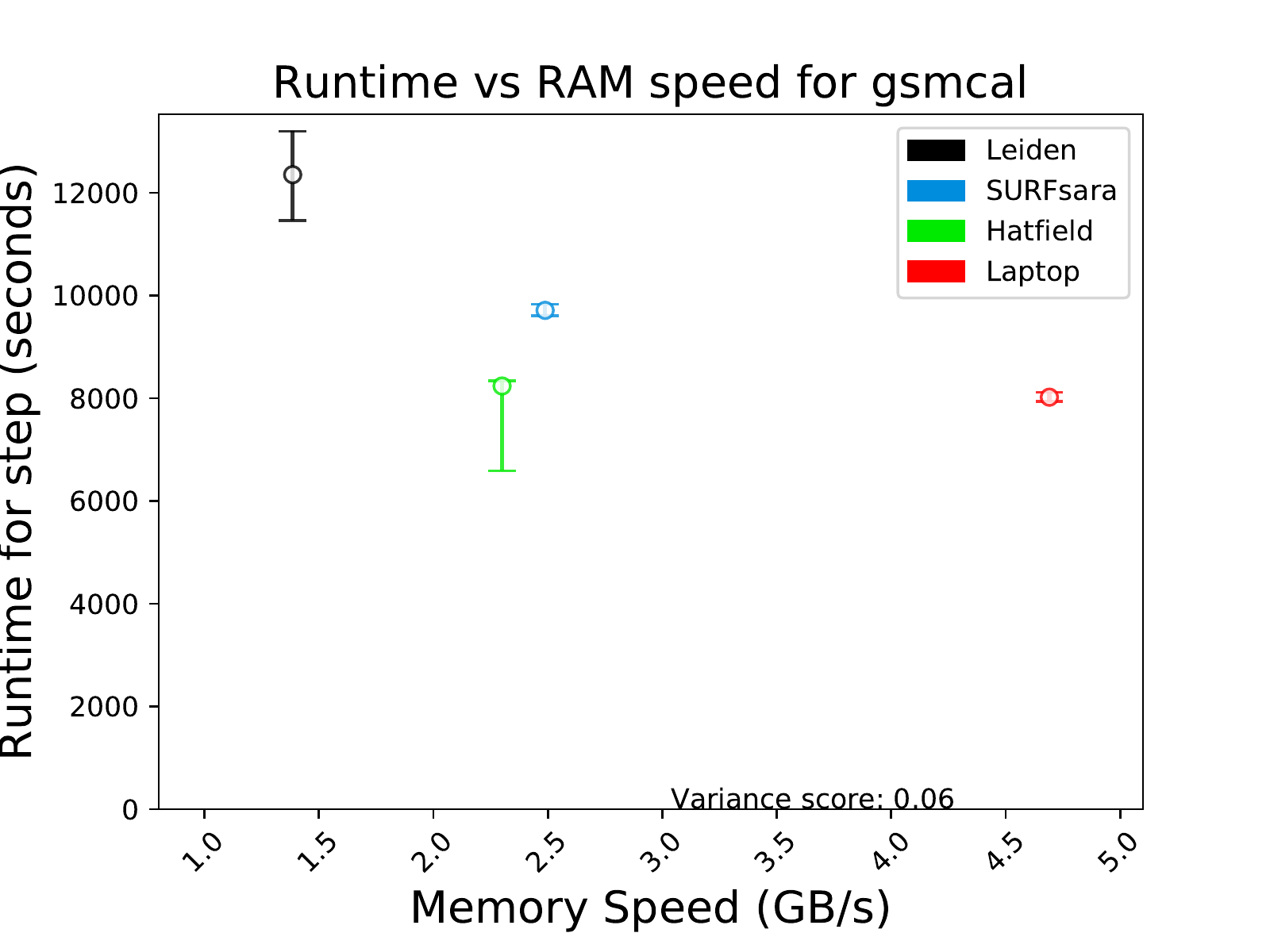}
      \caption{\textit{gsmcal\_solve}}
	\label{gamcal_RAM}
 \end{subfigure}
 \label{ram_bandw}
 \caption{Performance of the two bottleneck steps and RAM bandwidth in GB/s. Both the \textit{calib\_cal} and \textit{gsmcal\_solve} steps show a trend of faster processing times on machines with higher RAM bandwidth. Both steps show a trend of decreasing processing time with increasing RAM throughput.} 
\end{minipage}
\end{figure}

Figure \ref{calib_cal_RAM} showed that higher bandwidth is correlated with a faster completion time for the \textit{calib\_cal} and \textit{gsmcal\_solve} steps (\textbf{R4}). The result is to be expected as the working set of these steps is 200MB and 1.0GB respectively, and cannot fit into cache readily, however it is loaded into RAM within the first 5 seconds of the run (Figure \ref{fig:VMRSS}), and is streamed from memory throughout the run. 

\subsubsection{Disk Read speeds}

\begin{figure}
  \centering
   \begin{subfigure}{.45\textwidth}
    \includegraphics[width=\textwidth]{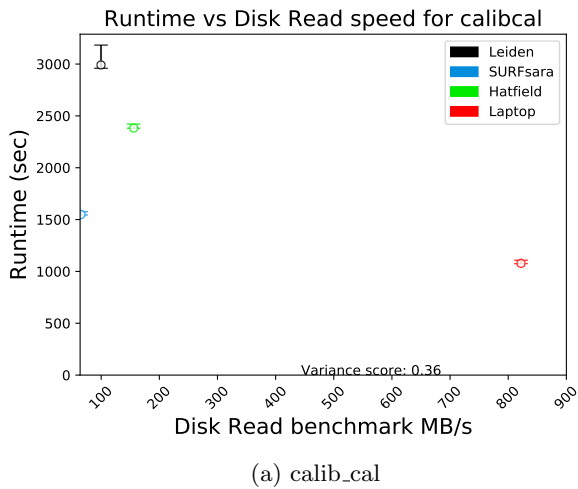}
      \caption{calib\_cal }
	\label{calib_cal_HDD}
 \end{subfigure}
  \begin{subfigure}{.45\textwidth}
    \includegraphics[width=\textwidth]{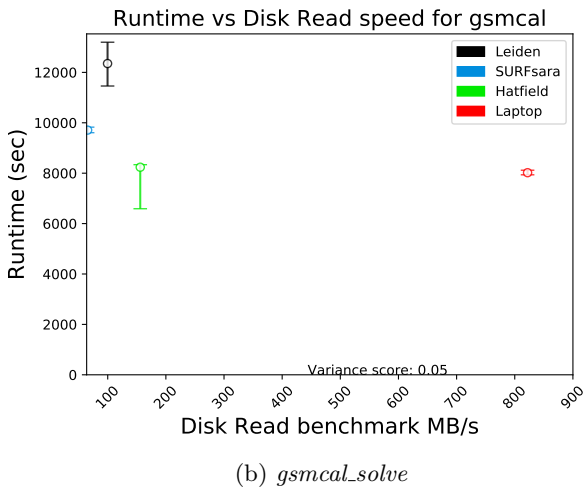}
      \caption{\textit{gsmcal\_solve}}
	\label{gsmcal_HDD}
 \end{subfigure}
 \label{fig:HDD_2_steps}
 \caption{Performance of the two bottleneck steps and Disk bandwidth in MB/s. There is no correlation between the Disk read speed and the Runtime of the steps.} 
\end{figure}

The slowest link in the memory hierarchy is the disk read speed. For the \textit{calib\_cal} step, the entire data is loaded into memory during the first few seconds of the run, after which the disk only becomes important when the results need to be written out. The \textit{gsmcal\_solve} step streams data from the disk to memory throughout the entire run.  The plot of disk read speeds (Fig. \ref{gsmcal_HDD}) also shows that a faster disk does not speed up the slowest step \textbf{R5}. To verify that disk throughput was not the limiting factor, the entire dataset (25 GB) was moved to main memory (using /dev/shm).  The resulting runtime for both bottleneck steps did not change.  

\begin{figure*}
  \centering
   \begin{subfigure}{.45\textwidth}
    \includegraphics[width=\textwidth]{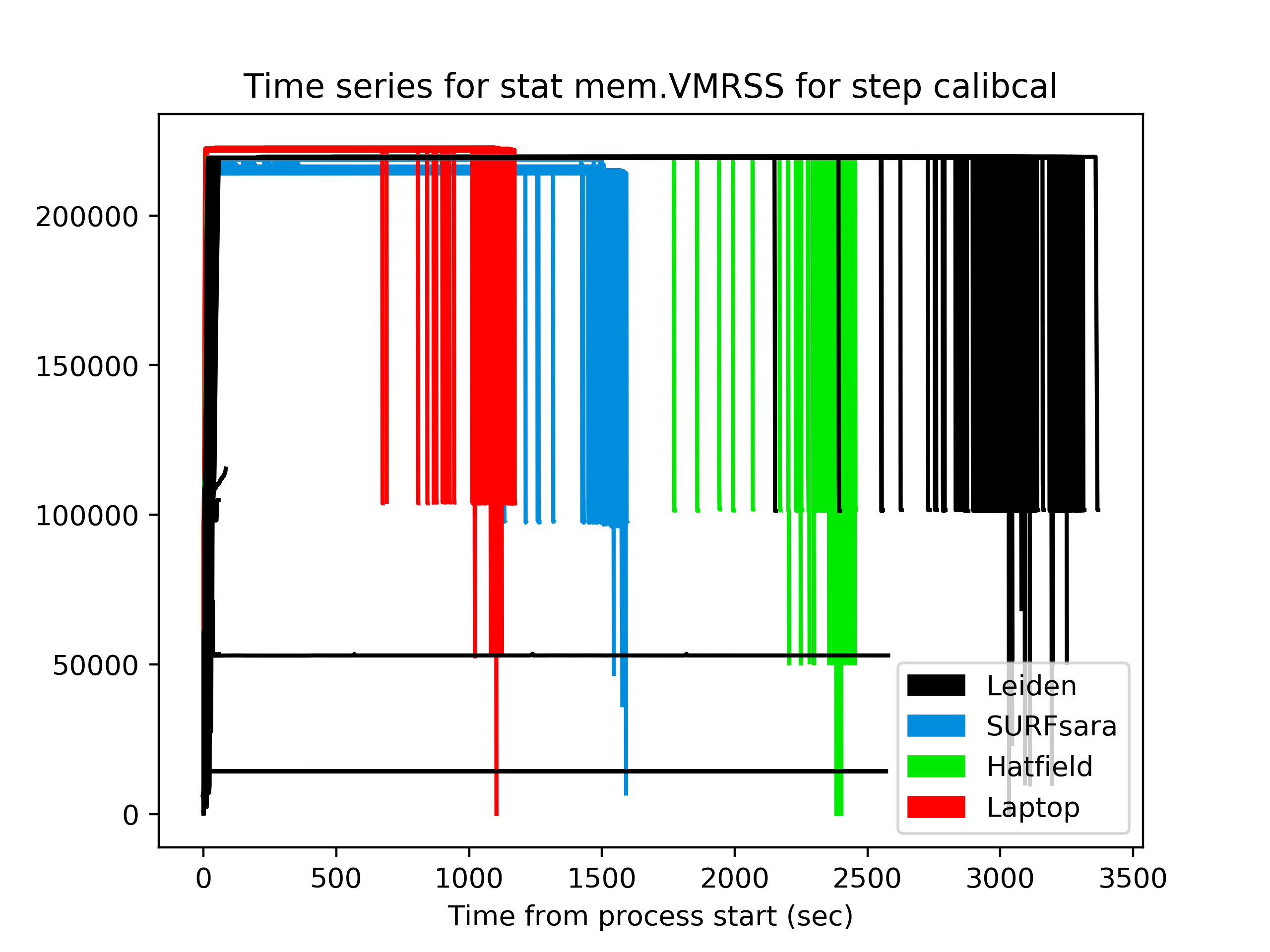}
      \caption{\textit{calib\_cal} }
	\label{calib_cal_VMRSS}
 \end{subfigure}%
 \begin{subfigure}{.45\textwidth}
    \includegraphics[width=\textwidth]{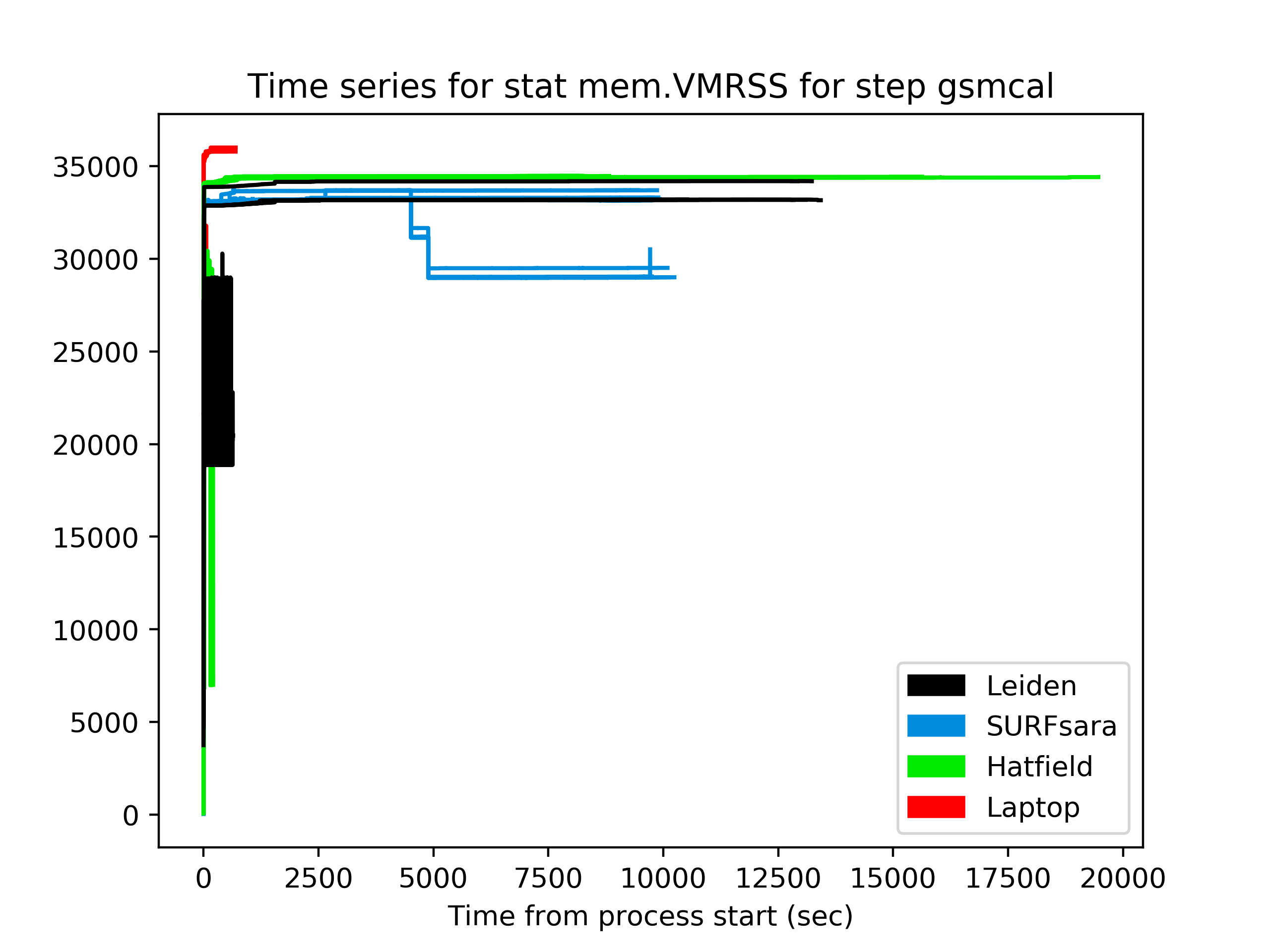}
      \caption{\textit{gsmcal\_solve}}
	\label{ffig:fitclock_VMRSS}
 \end{subfigure}
 \caption{Time series of the Virtual Memory Resident Set Size . This is the amount of data stored in RAM (in kB) during the \textit{calib\_cal} and \textit{gsmcal\_solve} steps. Both steps show the same amount of memory use on all test machines. Additionally, after a brief loading of data, the memory usage remains constant until processing is finished. }
 \label{fig:VMRSS}
\end{figure*}

The calibration steps both stored less than 200MB of data in memory throughout their run. Figure \ref{fig:VMRSS} shows the time-series of the total memory used by these steps. The \textit{calib\_cal} step uses only 200MB of memory and \textit{gsmcal\_solve} only 35MB. While the \textit{gsmcal\_solve} step works on a 1GB dataset, it streams the data in memory and thus does not require 1GB of RAM. Alternatively, the \textit{calib\_cal} step loads the entire (200MB) dataset into memory for the entire duration of the run. The RAM usage time-series in Figure \ref{fig:VMRSS} show that the RAM is filled for the first 5 seconds of the run, further confirming that the processing is effectively independent from disk speed.

\section{CPU Utilization Tests with PAPI}\label{sec:PAPI}

To gain more fine grained data on the CPU utilization, the \textit{calib\_cal} and \textit{gsmcal\_solve} steps were tested with the PAPI package.  We ran this package as a test, to determine whether collecting PAPI data is helpful in understanding pipeline performance.  PAPI can record data such as cache performance, branch prediction rate, fraction of memory/branch instructions and others. This data is complementary to the procfs information, which is collected by the Linux kernel. As the collected data was useful in understanding the \textit{prefactor} pipeline, we will include PAPI in the \textit{pipeline\_collector} suite in the future. In the following sections we will discuss the results obtained for the \textit{calib\_cal} and \textit{gsmcal\_solve} steps.

\subsection{Level 1 Data Misses}

\begin{figure}
  \centering
      \begin{subfigure}{.45\textwidth}
      \includegraphics[width=\textwidth]{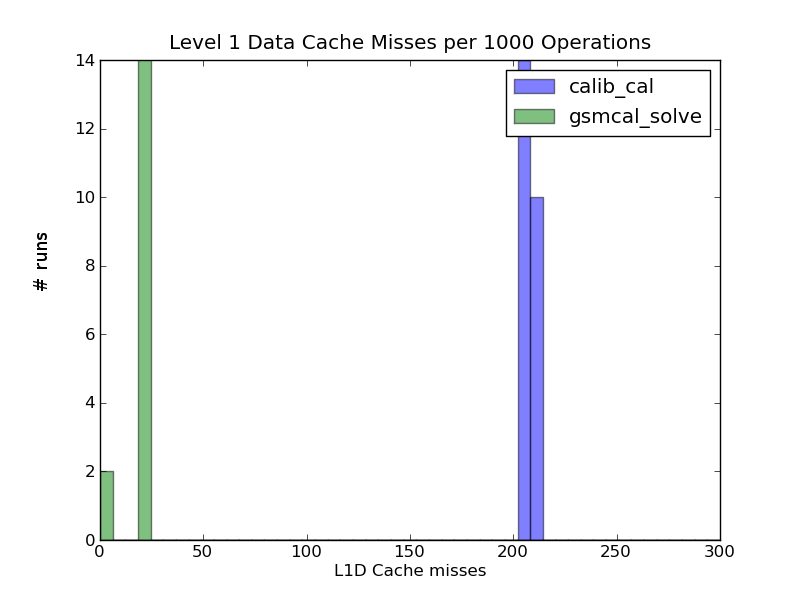}
      \caption{Level 1 Data cache misses }
	\label{fig:L1Dm}
    \end{subfigure}
    \begin{subfigure}{.45\textwidth}
    \includegraphics[width=\textwidth]{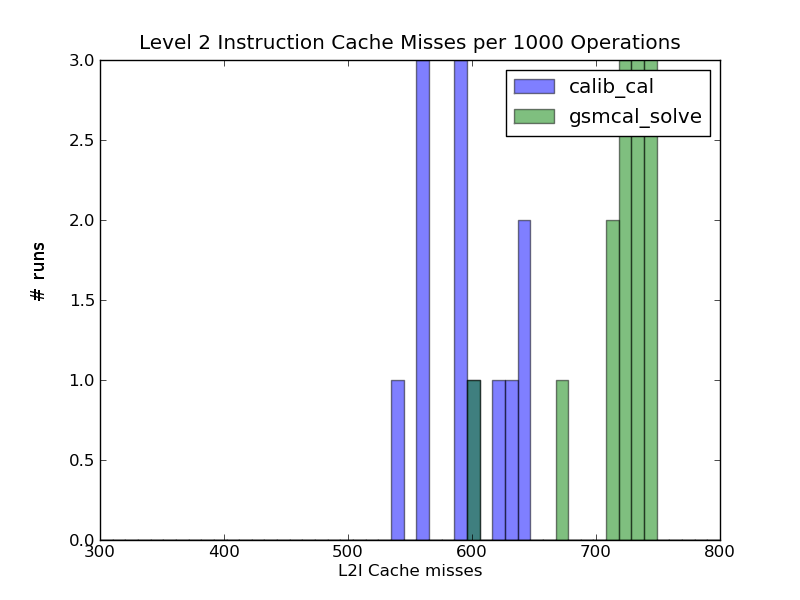}
      \caption{Level 2 Instruction cache misses }
	\label{fig:L2Im}
  \end{subfigure}
  \caption{Cache miss rates for \textit{calib\_cal} and \textit{gsmcal\_solve}, executed on the SURFsara gina cluster. The cache is split into instruction and data caches. The figures above show the difference in number of cache misses for both instruction cache and data cache for the slowest \textit{prefactor} steps. \textit{calib\_cal} suffers significantly more Data cache misses than \textit{gsmcal\_solve} while the two steps undergo similar instruction Cache misses.  }
\end{figure}

The Level 1 Cache is split into cache for instructions and data. For all our test hardware the L1 Data cache is 32 kB, and has a direct link to the processor's computational units \citep{haswell}. The processor collects information logging how many times data requested by the CPU is not located into the L1 Data cache. This counter is called the Level 1 Data Cache Miss rate. To resolve this type of cache miss, the data needs to be fetched from L2 Cache. When this happens, the processor has to wait for the requested data. \textbf{R7}: The recorded L1 data misses in Figure \ref{fig:L1Dm}, show that the software performing the \textit{calib\_cal} step  misses 20\% of its L1 data cache requests, while the software implementing the \textit{gsmcal\_solve} step misses less than 5\% of L1 Cache requests. These cache misses often happens in multi-threaded applications where there are instructions shared by multiple threads on the same cache line \citep{cache_opt}. 

\subsection{Level 2 Instruction Misses}

Unlike the Level 1 cache, Level 2  cache stores data and instructions in the same location. When the cache is full, it evicts the last used element in order to make space for newly requested data. PAPI also counts these eviction events. 
Figure \ref{fig:L2Im} shows that for both steps, between 50 and 70\% of L2 requests for an instruction do not match the contents of L2 Cache. This is significantly more than the applications benchmarked in \citep[Table 2]{cache_misses}. Because both steps process data of considerable size, the large amount of data required can evict instructions from the L2 cache (insight number \textbf{R7} in table \ref{table:results}). 

\subsection{Resource Stalls}

\begin{figure}
  \centering
      \begin{subfigure}{.45\textwidth}
      \includegraphics[width=\textwidth]{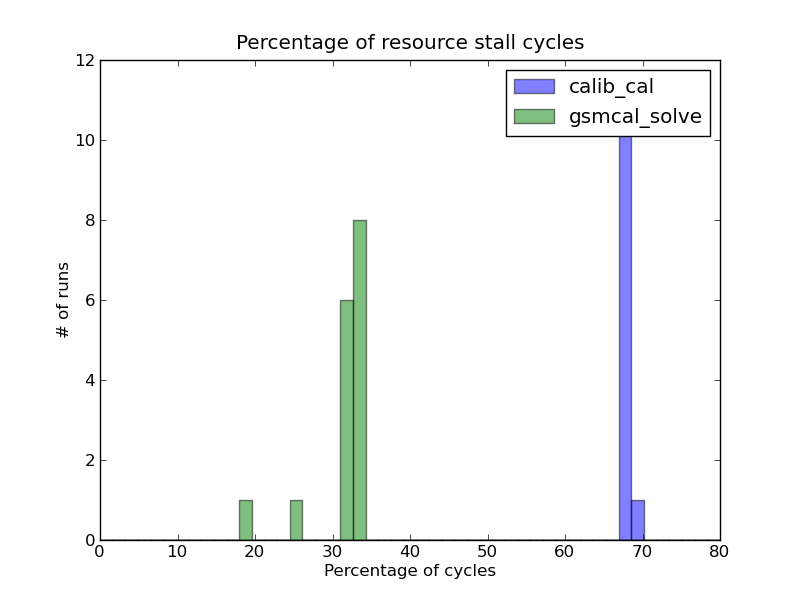}
      \caption{Resource Stall Cycles}
	\label{fig:rstall}
    \end{subfigure}
    \begin{subfigure}{.45\textwidth}
    \includegraphics[width=\textwidth]{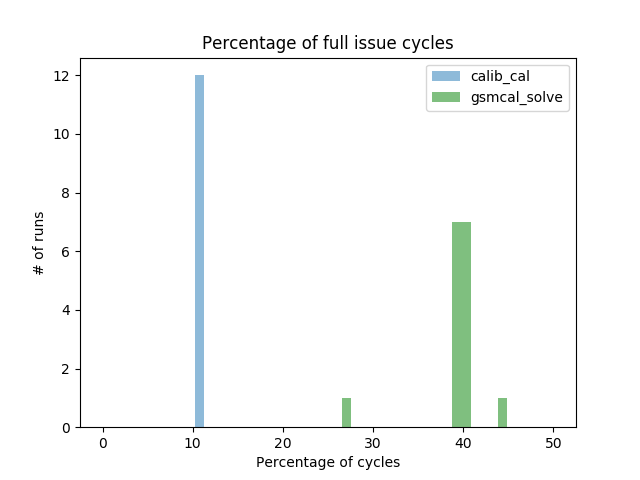}
      \caption{Full Issue cycles }
	\label{fig:full_issue}
  \end{subfigure}
  \label{fig:rr}
  \caption{Resource stall cycles and Full Instruction Issue cycles. The two steps were  executed on the SURFsara gina cluster. }
\end{figure}

Modern processors have multiple computational pipelines on chip, in order to process data in parallel \citep{pipeline_x86}. There are times when the processor's internal pipeline needs to wait for other instructions to finish. When this happens, it flags that it has 'stalled on a resource'. These resource stall cycles are also recorded by PAPI and represented as a percentage of total cycles. From figure \ref{fig:rstall}, it can be seen that \textit{calib\_cal} stalls on 70\% of the processor cycles, while  \textit{gsmcal\_solve} only on 33\% of cycles (\textbf{R8}). 

The Full Issue Cycles counter indicates the percentage of processor cycles, in which the theoretical maximum number of instructions are executed. During these cycles, the software uses the CPU optimally. The full issue cycles counter (Fig. \ref{fig:full_issue}) also shows the difference in efficiency between the \textit{calib\_cal} and \textit{gsmcal\_solve} step  (\textbf{R9}), with the former only working at peak efficiency for 10\% of the processor cycles.

The plots in Figures \ref{fig:rstall} and \ref{fig:full_issue} indicate that the \textit{calib\_cal} step does not use the internal CPU pipelines efficiently leading to waiting on resources and sub-optimal use of the CPU's Computational Units.

\section{Discussions and Recommendations}

With an increase of data acquisition rates and data complexity in radio astronomy, it is becoming important to thoroughly understand and optimize the performance of processing pipelines. Using \textit{pipeline\_collector}, data can be collected for each pipeline step without altering the processing software. We store this data in a time-series database. The collected data can be studied to help researchers understand the pipeline performance for different processing parameters, datasets, and on different hardware. The \textit{pipeline\_collector} suite is easy to deploy for mature pipelines and has minimal impact on pipeline performance. Typical CPU usage is $<$0.2\% with a memory footprint of $\sim$ 1-10 MB.

Creating a performance model with the collected data will allow us to to optimize future clusters for LOFAR data processing. Doing so is necessary given the current data throughput, number of observations and time-line of the SKSP project. Similar issues will be encountered with upcoming radio telescopes \citep{meerkat_ska_size}.

To showcase the power of the pipeline\_collector suite, the LOFAR \textit{prefactor} pipeline was run through a single data set on three clusters and a personal machine. A number of insights were made using the high resolution timing data collected from this package (such as in Figure \ref{fig:VMRSS}) and are listed in Table \ref{table:results}. In the future, we'll apply the \textit{pipeline\_collector} software to the more complex LOFAR DD pipeline, \textit{ddf-pipeline}\footnote{https://github.com/mhardcastle/ddf-pipeline}. 

The slowest processing steps for the \textit{prefactor} pipeline were identified as the \textit{calib\_cal} and \textit{gsmcal\_solve} steps. While the data can fit into the RAM for all of the processing machines, it is much larger than the processor's internal cache (Figure \ref{fig:mem_hiearch}).  The discoveries made concerned the memory hierarchy in Figure \ref{fig:mem_hiearch}. Results labeled \textbf{R2}, \textbf{R8} and \textbf{R9} related to the CPU performance; \textbf{R2}, \textbf{R6} and \textbf{R7} related to the Cache performance; \textbf{R3} and \textbf{R5} related to the Memory usage and \textbf{R4} discussed the Disk speed. 

Faster processors did not accelerate the \textit{gsmcal\_solve} step significantly, as this step streams data between the RAM and CPU. As the CPU speed increases, streaming applications become bottlenecked by the throughput of data into the CPU from RAM. As the \textit{gsmcal\_solve} algorithm iteratively calibrates chunks of the data, these chunks need to be loaded from disk once, however they are moved from RAM to CPU multiple times during calibration. 

Similarly, the \textit{calib\_cal} step is more dependent on memory throughput than on CPU speed as this step moves data to and from memory frequently. This step also does minimization looping over the dataset. As the dataset does not fit in the cache, parts of it need to be constantly moving from memory and back. Figure \ref{calib_cal_cache} shows that the machine with the smallest LLC cache runs the \textit{calib\_cal} step the fastest. This is likely a combination of the benefit of faster RAM and poor cache optimization for this software. The same effect is much less pronounced in Figure \ref{gsmcal_cache}, suggesting that software optimization at least plays a part in the outliers for the laptop machine. 

\subsection{Recommendations}
Based on these results, the top hardware recommendation is that \textit{prefactor}'s slowest steps can be accelerated by running on machines with faster memory or upgrading the memory of the current machines. The two slowest \textit{prefactor} steps showed improvements on machines with faster RAM.  

One software recommendation is to improve the efficiency of the \textit{calib\_cal} step through refactoring or by replacing the software package used. Unfortunately, the software used for the \textit{gsmcal\_solve} step cannot be used for the \textit{calib\_cal} step as it is not yet able to  correct for Faraday Rotation \citep{stefcal}, making it impossible to currently use the software used by the \textit{gsmcal\_solve} step. Faraday Rotation has recently been implemented in a development version of the \textit{prefactor} pipeline and is currently undergoing testing. This version of the pipeline will be implemented by September 2018.

Additionally, the large number of data cache misses recorded for the \textit{calib\_cal} step suggests that its source code is not optimized for multi-threaded processing. Data cache misses are often encountered when multiple threads have instructions on the same cache line\footnote{A cache line is a row of cache memory which is loaded into CPU as a single unit \citep{cache_architecture} }, forcing the memory controller to move this cache line between cores \citep{cache_misses}. This can also explain the large number of stalled cycles (Fig. \ref{fig:rstall}) and low number of full issue cycles (Fig. \ref{fig:full_issue}) for the \textit{calib\_cal} step. It is recommended to further study the inefficiencies of  \textit{calib\_cal} or to replace it with a newer software. If the software processing for this step is updated, analysing the cache and CPU performance of the new software will be necessary to determine whether it efficiently uses the available computational resources.

Finally, we discovered that compiling the software on a virtual machine did not lead to a processing slowdown. This means that the current slowest \textit{prefactor} steps are not optimized to use advanced processor instructions. Nevertheless, the resulting cross-compatibility is an encouraging result as it will allow to easily distribute pre-compiled versions of the software without increasing the processing time. We recommend continuing CVMFS deployment of LOFAR software.

\section{Conclusions}
 
In this paper, we present a novel system for automated collection of performance data for complex software pipelines. We use this suite to study the LOFAR \textit{prefactor} pipeline. The results are discussed aiming to understand the effect of different hardware parameters on the data processing. To do so, we run the pipeline on four different machines. 

The software automatically collects performance data at the operating system level without impacting processing time. Data for each pipeline step is extracted using the OpenTSDB API, plotted and analyzed. Additionally, the \textit{pipeline\_collector} suite is easy to extend with new collectors that record more detailed time-series data for each pipeline step. The performance data is stored in the time series database OpenTSDB.

Here, we used this data to find 9 insights into the LOFAR prefactor pipeline listed in Table \ref{table:results}. The implementation details are described in \ref{appendix1}.

The \textit{prefactor} pipeline is used to do the initial processing for over 3000 observations that are part of the LOFAR SKSP Tier 1 survey. However, this pipeline is also used for lots of other LOFAR datasets outside the SKSP project. We have shown that increasing the RAM throughput is the easiest way to speedup \textit{prefactor} processing. Running the \textit{calib\_cal} step on hardware with RAM faster than 4 GB/s will save up to 700k CPU hours for the 3000+ unprocessed datasets. This throughput increase will also speedup the \textit{gsmcal\_solve} step by 30\% saving an additional 400k CPU hours. This is a significant fraction of the estimated 2,400k CPU hours required to process this data with the \textit{prefactor} pipeline. 

As shown in this work, we can correlate the performance of the LOFAR software with different hardware specifications. Additionally, the datasets  can vary in size and job overheads on the compute cluster can depend on the processing parameters. All of these parameters affect the processing latency for the calibration and imaging pipelines. As such, a thorough parametric model is required to further optimize the end-to-end LOFAR processing pipeline and predict processing times on future clusters.

The design of this utility makes it easy to apply to future LOFAR pipelines. It is important to note that \textit{pipeline\_collector} is general enough that it can be used by other scientific pipelines, with no modification of the pipeline. Integrating \textit{pipeline\_collector} with a different pipeline requires only minor work. In future work, we will integrate \textit{pipeline\_collector} with \textit{ddf-pipeline}. We will use this data to create a performance model of the full LOFAR imaging pipeline \citep{lofar_prefactor,Wendy_bootes,tassesmirnov}.  including the DI step, implemented by \textit{prefactor}, and  the DD step, implemented by \textit{ddf-pipeline}. This model will make it possible to first understand and then optimize the LOFAR pipeline and suggest for hardware and software improvements. 

\appendix

\section{Performance Collection Implementation Details}\label{appendix1}

In this work, we have developed the \textit{pipeline\_collector} suite, aimed at collecting detailed time-series information from distributed scientific pipelines. 

The \textit{tcollector} package is a python software suite that can collect system performance data at predetermined intervals. The package is designed to monitor the performance statistics for web-servers and cluster nodes. The tcollector software records time series of the different performance metrics and sends them to a Time Series Database through HTTP. The Time Series Database, OpenTSDB stores the data in an HBase \citep{hbase} instance at the performance collection server. Users interested in plotting time series can plot real time or historical data through an HTTP interface with OpenTSDB. With a central performance collection server, data from multiple processing sites can be collected and analyzed. 

\begin{figure}[!h]
    \includegraphics[width=\linewidth]{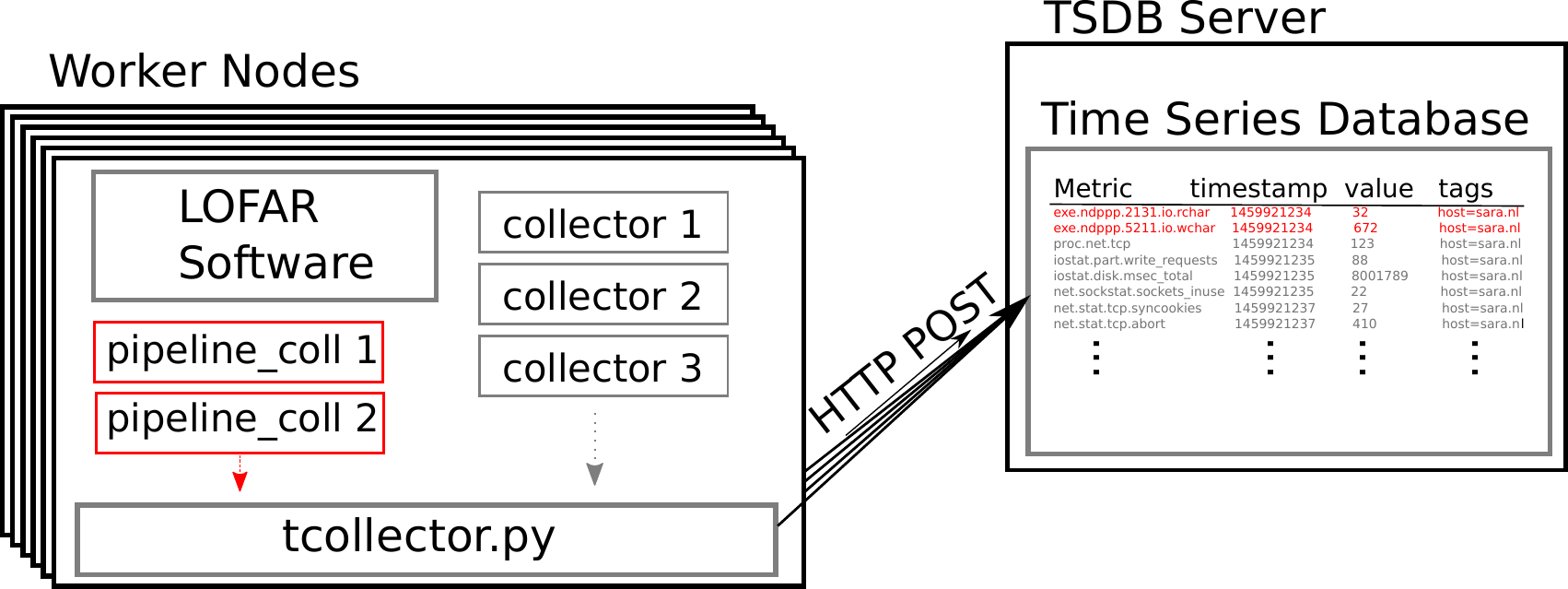}
      \caption{Communication between worker nodes and the TSDB server, including the pipeline\_collector modules (in red). The \textit{pipeline\_collector} suite collects information on the running LOFAR pipelines, while the rest of the tcollector package collects system performance data. The existing tcollector package and its collectors are shown in gray. The collectors in gray only record metrics from the global system. }
	\label{fig:tsdb_tcollector}
\end{figure}

Tcollector formats the time series information in four fields. First is the name of the metric which is measured. Second is the UNIX timestamp. Third is the time series recorded as an integer or a float. Finally, a set of tags (key-value pairs) can be added to the data point. These four fields are discussed below and can be seen on the right side of Figure \ref{fig:tsdb_tcollector}. 

\subsection{The pipeline\_collector suite}\label{sec:customcollectors}

The tcollector package cannot collect data on individual processes, nor can it associate these processes with specific steps of a data processing pipeline. We've supplemented the software with the \textit{pipeline\_collector} suite\footnote{Located at https://gitlab.com/apmechev/procfs\_tcollector.git} using an executable that monitors a pipeline's running processes\footnote{Located at https://github.com/apmechev/procfsamp}. When an executable that is part of the LOFAR pipeline launches, a dedicated collector begins reporting information on the individual process. Every time a new processing step starts, the \textit{prefactor} pipeline records the step name in a log file. \textit{pipeline\_collector} determines the current running step using this log. Running the LOFAR processing concurrently with the tcollector package gives us per-step performance data without changing or slowing down the LOFAR \textit{prefactor} pipeline. Furthermore, \textit{pipelie\_collector} can be integrated with any processing pipeline as long as each pipeline step's name is recorded at its launch. 

The \textit{pipeline\_collector} suite sends data to the time series database in the same format as the rest of the collectors included in the tcollector package. 

\subsection{Setting up for future pipelines}

The setup options for \textit{pipeline\_collector} are stored in a configuration file in the root directory of the package. This file holds the sample interval, executables to monitor and the location where \textit{pipeline\_collector} can read the current pipeline step

The \textit{pipeline\_collector} suite reads the current pipeline step from a file, the location of which is specified in the configuration. This file needs to be updated each time the pipeline begins a new step. For LOFAR we have a script running with the pipeline, and determining the current step using the pipeline logs. As each pipeline has a unique sequence of steps, the current step needs to be recorded in a file in order for \textit{pipeline\_collector} to report it to the time series database. The location of the file recording the current pipeline step is read from the configuration file. 

Next, the names of the specific processes need to be included in the configuration file. In the case of LOFAR, we select the \texttt{NDPPP}, \texttt{bbs-reducer} and \texttt{losoto} processes. The \textit{pipeline\_collector} searches the running processes for the current user for these process names and launches a collector for each new process launched by the current step.

\section*{Acknowledgements}
APM would like to acknowledge the support from the NWO/DOME/IBM programme ``Big Bang Big Data: Innovating ICT as a Driver For Astronomy'', project \#628.002.001.

HJR gratefully acknowledge support from the European Research Council under the European Unions Seventh Framework Programme (FP/2007-2013)/ERC Advanced Grant NEWCLUSTERS- 321271.

JBRO acknowledges financial support from NWO Top LOFAR-CRRL project, project No. 614.001.351.

This work was carried out on the Dutch national e-infrastructure with the support of SURF
Cooperative through grant e-infra 160022 \& 160152.

\bibliography{bibliography}

\begin{thebibliography}{44}
\providecommand{\natexlab}[1]{#1}
\providecommand{\url}[1]{\texttt{#1}}
\expandafter\ifx\csname urlstyle\endcsname\relax
  \providecommand{\doi}[1]{doi: #1}\else
  \providecommand{\doi}{doi: \begingroup \urlstyle{rm}\Url}\fi

\bibitem[Ahn(2008)]{papiex}
D.~H. Ahn.
\newblock Measuring flops using hardware performance counter technologies on lc
  systems.
\newblock Technical report, Lawrence Livermore National Laboratory (LLNL),
  Livermore, CA, 2008.

\bibitem[Apache(2017)]{tcollector}
S.~F. Apache.
\newblock Tcollector: {OpenTSDB} documentation.
\newblock Available at
  \url{http://opentsdb.net/docs/build/html/user_guide/utilities/tcollector.html
  }, 2017.

\bibitem[Apache~HBase(2015)]{hbase}
T.~Apache~HBase.
\newblock Apache hbase reference guide.
\newblock \emph{Apache, version}, 2\penalty0 (0), 2015.

\bibitem[Blomer et~al.(2011)Blomer, Aguado-S{\'a}nchez, Buncic, and
  Harutyunyan]{cvmfs}
J.~Blomer, C.~Aguado-S{\'a}nchez, P.~Buncic, and A.~Harutyunyan.
\newblock Distributing {LHC} application software and conditions databases
  using the cernvm file system.
\newblock In \emph{Journal of Physics: Conference Series}, volume 331, page
  042003. IOP Publishing, 2011.

\bibitem[Bowden(2009)]{procfs}
T.~Bowden.
\newblock {THE} /proc {FILESYSTEM} v1.3.
\newblock \url{https://www.kernel.org/doc/Documentation/filesystems/proc.txt},
  2009.

\bibitem[Broekema et~al.(2015)Broekema, van Nieuwpoort, and
  Bal]{meerkat_ska_size}
P.~C. Broekema, R.~V. van Nieuwpoort, and H.~E. Bal.
\newblock The {S}quare {K}ilometre {A}rray science data processor.
  {P}reliminary compute platform design.
\newblock \emph{Journal of Instrumentation}, 10\penalty0 (07):\penalty0 C07004,
  2015.

\bibitem[Broekema et~al.(2018)Broekema, Mol, Nijboer, van Amesfoort, Brentjens,
  Loose, Klijn, and Romein]{lofarcobalt}
P.~C. Broekema, J.~J.~D. Mol, R.~Nijboer, A.~van Amesfoort, M.~Brentjens, G.~M.
  Loose, W.~Klijn, and J.~Romein.
\newblock Cobalt: A gpu-based correlator and beamformer for lofar.
\newblock \emph{Astronomy and Computing}, 23:\penalty0 180 -- 192, 2018.
\newblock ISSN 2213-1337.
\newblock \doi{https://doi.org/10.1016/j.ascom.2018.04.006}.
\newblock URL
  \url{http://www.sciencedirect.com/science/article/pii/S2213133717301439}.

\bibitem[Centeno et~al.(2014)Centeno, Schou, Hayashi, Norton, Hoeksema, Liu,
  Leka, and Barnes]{solar_pipeline}
R.~Centeno, J.~Schou, K.~Hayashi, A.~Norton, J.~Hoeksema, Y.~Liu, K.~Leka, and
  G.~Barnes.
\newblock The {H}elioseismic and {M}agnetic {I}mager ({H}{M}{I}) vector
  magnetic field pipeline: optimization of the spectral line inversion code.
\newblock \emph{Solar Physics}, 289\penalty0 (9):\penalty0 3531--3547, 2014.

\bibitem[David and John(2005)]{cache_architecture}
A.~P. David and L.~H. John.
\newblock Computer organization and design: the hardware/software interface.
\newblock \emph{San mateo, CA: M organ Kaufmann Publishers}, 1:\penalty0 998,
  2005.

\bibitem[Davidson(2012)]{meerkat_size}
D.~B. Davidson.
\newblock Potential technological spin-offs from {M}eer{K}{A}{T} and the
  {S}outh {A}frican {S}quare {K}ilometre {A}rray bid.
\newblock \emph{South African Journal of Science}, 108\penalty0 (1-2):\penalty0
  01--03, 2012.

\bibitem[Denning(1968)]{workingset}
P.~J. Denning.
\newblock The working set model for program behavior.
\newblock \emph{Communications of the ACM}, 11\penalty0 (5):\penalty0 323--333,
  1968.

\bibitem[Dijkema(2017)]{cookbook}
T.~J. Dijkema.
\newblock {LOFAR} {I}maging {C}ookbook.
\newblock Available at
  \url{http://www.astron.nl/sites/astron.nl/files/cms/lofar_imaging_cookbook_v19.pdf
  }, 2017.

\bibitem[Goto and Geijn(2008)]{goto2008anatomy}
K.~Goto and R.~A. Geijn.
\newblock Anatomy of high-performance matrix multiplication.
\newblock \emph{ACM Transactions on Mathematical Software (TOMS)}, 34\penalty0
  (3):\penalty0 12, 2008.

\bibitem[Gregg and Mauro(2011)]{dtrace}
B.~Gregg and J.~Mauro.
\newblock \emph{DTrace: Dynamic Tracing in Oracle Solaris, Mac OS X and
  FreeBSD}.
\newblock Prentice Hall Professional, 2011.

\bibitem[Gupta et~al.(2017)Gupta, Ajithkumar, Kale, Nayak, Sabhapathy,
  Sureshkumar, Swami, Chengalur, Ghosh, Ishwara-Chandra, et~al.]{gmrt_upgrade}
Y.~Gupta, B.~Ajithkumar, H.~Kale, S.~Nayak, S.~Sabhapathy, S.~Sureshkumar,
  R.~Swami, J.~Chengalur, S.~Ghosh, C.~Ishwara-Chandra, et~al.
\newblock The upgraded {G}{M}{R}{T}: opening new windows on the radio
  {U}niverse.
\newblock \emph{Current Science}, 113\penalty0 (4):\penalty0 707, 2017.

\bibitem[Hazelwood and Smith(2004)]{cache_eviction}
K.~Hazelwood and J.~E. Smith.
\newblock Exploring code cache eviction granularities in dynamic optimization
  systems.
\newblock In \emph{Code Generation and Optimization, 2004. CGO 2004.
  International Symposium on}, pages 89--99. IEEE, 2004.

\bibitem[Hu et~al.(2006)Hu, Kim, Lipasti, and Smith]{pipeline_x86}
S.~Hu, I.~Kim, M.~H. Lipasti, and J.~E. Smith.
\newblock An approach for implementing efficient superscalar {CISC} processors.
\newblock In \emph{High-Performance Computer Architecture, 2006. The Twelfth
  International Symposium on}, pages 41--52. IEEE, 2006.

\bibitem[Jain and Agrawal(2013)]{haswell}
T.~Jain and T.~Agrawal.
\newblock The haswell microarchitecture-4th generation processor.
\newblock \emph{International Journal of Computer Science and Information
  Technologies}, 4\penalty0 (3):\penalty0 477--480, 2013.

\bibitem[Jonas(2009)]{meerkat}
J.~L. Jonas.
\newblock Meer{K}{A}{T}-{T}he {S}outh {A}frican array with composite dishes and
  wide-band single pixel feeds.
\newblock \emph{Proceedings of the IEEE}, 97\penalty0 (8):\penalty0 1522--1530,
  2009.

\bibitem[Katz and Patterson(2001)]{mem_hiearch}
R.~H. Katz and D.~A. Patterson.
\newblock Memory hierarchy, {CMPUT429/CMPE382} {Winter} 2001. {University of
  Calgary}.
\newblock Available at
  \url{https://webdocs.cs.ualberta.ca/~amaral/courses/429/webslides/Topic4-MemoryHierarchy/sld003.htm
  }, 2001.

\bibitem[Lebeck et~al.(2002)Lebeck, Koppanalil, Li, Patwardhan, and
  Rotenberg]{cache_misses}
A.~R. Lebeck, J.~Koppanalil, T.~Li, J.~Patwardhan, and E.~Rotenberg.
\newblock A large, fast instruction window for tolerating cache misses.
\newblock In \emph{Computer Architecture, 2002. Proceedings. 29th Annual
  International Symposium on}, pages 59--70. IEEE, 2002.

\bibitem[Lonsdale et~al.(2009)Lonsdale, Cappallo, Morales, Briggs, Benkevitch,
  Bowman, Bunton, Burns, Corey, Doeleman, et~al.]{MWA}
C.~J. Lonsdale, R.~J. Cappallo, M.~F. Morales, F.~H. Briggs, L.~Benkevitch,
  J.~D. Bowman, J.~D. Bunton, S.~Burns, B.~E. Corey, S.~S. Doeleman, et~al.
\newblock The murchison widefield array: {D}esign overview.
\newblock \emph{Proceedings of the IEEE}, 97\penalty0 (8):\penalty0 1497--1506,
  2009.

\bibitem[Loose(2008)]{bbs_selfcal}
G.~Loose.
\newblock {LOFAR} self-calibration using a blackboard software architecture.
\newblock In \emph{Astronomical Data Analysis Software and Systems XVII},
  volume 394, page~91, 2008.

\bibitem[Massie et~al.(2004)Massie, Chun, and Culler]{ganglia}
M.~L. Massie, B.~N. Chun, and D.~E. Culler.
\newblock The ganglia distributed monitoring system: design, implementation,
  and experience.
\newblock \emph{Parallel Computing}, 30\penalty0 (7):\penalty0 817--840, 2004.

\bibitem[{Mechev} et~al.(2017){Mechev}, {Oonk}, {Danezi}, {Shimwell},
  {Schrijvers}, {Intema}, {Plaat}, and {Rottgering}]{mechev}
A.~{Mechev}, J.~B.~R. {Oonk}, A.~{Danezi}, T.~W. {Shimwell}, C.~{Schrijvers},
  H.~{Intema}, A.~{Plaat}, and H.~J.~A. {Rottgering}.
\newblock {An {A}utomated {S}calable {F}ramework for {D}istributing {R}adio
  {A}stronomy {P}rocessing {A}cross {C}lusters and {C}louds}.
\newblock In \emph{Proceedings of the International Symposium on Grids and
  Clouds (ISGC) 2017, held 5-10 March, 2017 at Academia Sinica, Taipei, Taiwan
  (ISGC2017). Online at
  \url{https://pos.sissa.it/cgi-bin/reader/conf.cgi?confid=293}, id.2}, page~2,
  Mar. 2017.

\bibitem[Mucci et~al.(1999)Mucci, Browne, Deane, and Ho]{papi}
P.~J. Mucci, S.~Browne, C.~Deane, and G.~Ho.
\newblock {PAPI}: A portable interface to hardware performance counters.
\newblock In \emph{Proceedings of the department of defense HPCMP users group
  conference}, volume 710, 1999.

\bibitem[Nethercote and Seward(2007)]{valgrind}
N.~Nethercote and J.~Seward.
\newblock Valgrind: a framework for heavyweight dynamic binary instrumentation.
\newblock In \emph{ACM Sigplan notices}, volume~42, pages 89--100. ACM, 2007.

\bibitem[Offringa et~al.(2013)Offringa, De~Bruyn, Zaroubi, van Diepen,
  Martinez-Ruby, Labropoulos, Brentjens, Ciardi, Daiboo, Harker,
  et~al.]{lofar_NDPPP}
A.~Offringa, A.~De~Bruyn, S.~Zaroubi, G.~van Diepen, O.~Martinez-Ruby,
  P.~Labropoulos, M.~A. Brentjens, B.~Ciardi, S.~Daiboo, G.~Harker, et~al.
\newblock The {LOFAR} radio environment.
\newblock \emph{Astronomy \& astrophysics}, 549:\penalty0 A11, 2013.

\bibitem[{Oonk} et~al.(2014){Oonk}, {van Weeren}, {Salgado}, {Morabito},
  {Tielens}, {Rottgering}, {Asgekar}, {White}, {Alexov}, {Anderson}, {Avruch},
  {Batejat}, {Beck}, {Bell}, {van Bemmel}, {Bentum}, {Bernardi}, {Best},
  {Bonafede}, {Breitling}, {Brentjens}, {Broderick}, {Br{\"u}ggen}, {Butcher},
  {Ciardi}, {Conway}, {Corstanje}, {de Gasperin}, {de Geus}, {de Vos},
  {Duscha}, {Eisl{\"o}ffel}, {Engels}, {van Enst}, {Falcke}, {Fallows},
  {Fender}, {Ferrari}, {Frieswijk}, {Garrett}, {Grie{\ss}meier}, {Hamaker},
  {Hassall}, {Heald}, {Hessels}, {Hoeft}, {Horneffer}, {van der Horst},
  {Iacobelli}, {Jackson}, {Juette}, {Karastergiou}, {Klijn}, {Kohler},
  {Kondratiev}, {Kramer}, {Kuniyoshi}, {Kuper}, {van Leeuwen}, {Maat},
  {Macario}, {Mann}, {Markoff}, {McKean}, {Mevius}, {Miller-Jones}, {Mol},
  {Mulcahy}, {Munk}, {Norden}, {Orru}, {Paas}, {Pandey-Pommier}, {Pandey},
  {Pizzo}, {Polatidis}, {Reich}, {Scaife}, {Schoenmakers}, {Schwarz},
  {Shulevski}, {Sluman}, {Smirnov}, {Sobey}, {Stappers}, {Steinmetz},
  {Swinbank}, {Tagger}, {Tang}, {Tasse}, {Veen}, {Thoudam}, {Toribio}, {van
  Nieuwpoort}, {Vermeulen}, {Vocks}, {Vogt}, {Wijers}, {Wise}, {Wucknitz},
  {Yatawatta}, {Zarka}, and {Zensus}]{oonk_2014}
J.~B.~R. {Oonk}, R.~J. {van Weeren}, F.~{Salgado}, L.~K. {Morabito},
  A.~G.~G.~M. {Tielens}, H.~J.~A. {Rottgering}, A.~{Asgekar}, G.~J. {White},
  A.~{Alexov}, J.~{Anderson}, I.~M. {Avruch}, F.~{Batejat}, R.~{Beck}, M.~E.
  {Bell}, I.~{van Bemmel}, M.~J. {Bentum}, G.~{Bernardi}, P.~{Best},
  A.~{Bonafede}, F.~{Breitling}, M.~{Brentjens}, J.~{Broderick},
  M.~{Br{\"u}ggen}, H.~R. {Butcher}, B.~{Ciardi}, J.~E. {Conway},
  A.~{Corstanje}, F.~{de Gasperin}, E.~{de Geus}, M.~{de Vos}, S.~{Duscha},
  J.~{Eisl{\"o}ffel}, D.~{Engels}, J.~{van Enst}, H.~{Falcke}, R.~A. {Fallows},
  R.~{Fender}, C.~{Ferrari}, W.~{Frieswijk}, M.~A. {Garrett},
  J.~{Grie{\ss}meier}, J.~P. {Hamaker}, T.~E. {Hassall}, G.~{Heald}, J.~W.~T.
  {Hessels}, M.~{Hoeft}, A.~{Horneffer}, A.~{van der Horst}, M.~{Iacobelli},
  N.~J. {Jackson}, E.~{Juette}, A.~{Karastergiou}, W.~{Klijn}, J.~{Kohler},
  V.~I. {Kondratiev}, M.~{Kramer}, M.~{Kuniyoshi}, G.~{Kuper}, J.~{van
  Leeuwen}, P.~{Maat}, G.~{Macario}, G.~{Mann}, S.~{Markoff}, J.~P. {McKean},
  M.~{Mevius}, J.~C.~A. {Miller-Jones}, J.~D. {Mol}, D.~D. {Mulcahy},
  H.~{Munk}, M.~J. {Norden}, E.~{Orru}, H.~{Paas}, M.~{Pandey-Pommier}, V.~N.
  {Pandey}, R.~{Pizzo}, A.~G. {Polatidis}, W.~{Reich}, A.~M.~M. {Scaife},
  A.~{Schoenmakers}, D.~{Schwarz}, A.~{Shulevski}, J.~{Sluman}, O.~{Smirnov},
  C.~{Sobey}, B.~W. {Stappers}, M.~{Steinmetz}, J.~{Swinbank}, M.~{Tagger},
  Y.~{Tang}, C.~{Tasse}, S.~t. {Veen}, S.~{Thoudam}, C.~{Toribio}, R.~{van
  Nieuwpoort}, R.~{Vermeulen}, C.~{Vocks}, C.~{Vogt}, R.~A.~M.~J. {Wijers},
  M.~W. {Wise}, O.~{Wucknitz}, S.~{Yatawatta}, P.~{Zarka}, and A.~{Zensus}.
\newblock {Discovery of carbon radio recombination lines in absorption towards
  {Cygnus} {A}}.
\newblock \emph{MNRAS}, 437:\penalty0 3506--3515, Feb. 2014.
\newblock \doi{10.1093/mnras/stt2158}.

\bibitem[{Ott}(2010)]{herschel}
S.~{Ott}.
\newblock {The Herschel Data Processing System {\textemdash} HIPE and Pipelines
  {\textemdash} Up and Running Since the Start of the Mission}.
\newblock In \emph{Astronomical Data Analysis Software and Systems XIX}, volume
  434, page 139, Dec. 2010.

\bibitem[Salvini and Wijnholds(2014)]{stefcal}
S.~Salvini and S.~J. Wijnholds.
\newblock St{E}{F}{C}al—{A}n {A}lternating {D}irection {I}mplicit method for
  fast full polarization array calibration.
\newblock In \emph{General Assembly and Scientific Symposium (URSI GASS), 2014
  XXXIth URSI}, pages 1--4. IEEE, 2014.

\bibitem[Sarkar and Tullsen(2008)]{cache_opt}
S.~Sarkar and D.~Tullsen.
\newblock Compiler techniques for reducing data cache miss rate on a
  multithreaded architecture.
\newblock \emph{High Performance Embedded Architectures and Compilers}, pages
  353--368, 2008.

\bibitem[Shimwell et~al.(2017)Shimwell, R{\"o}ttgering, Best, Williams,
  Dijkema, De~Gasperin, Hardcastle, Heald, Hoang, Horneffer, et~al.]{lotss}
T.~Shimwell, H.~R{\"o}ttgering, P.~N. Best, W.~Williams, T.~Dijkema,
  F.~De~Gasperin, M.~Hardcastle, G.~Heald, D.~Hoang, A.~Horneffer, et~al.
\newblock The {L}{O}{F}{A}{R} {T}wo-metre {S}ky {S}urvey-{I}. {S}urvey
  description and preliminary data release.
\newblock \emph{Astronomy \& Astrophysics}, 598:\penalty0 A104, 2017.

\bibitem[Sigoure(2012)]{opentsdbsite}
B.~Sigoure.
\newblock Open{T}{S}{D}{B} scalable time series database ({T}{S}{D}{B}), 2012.

\bibitem[Skadron et~al.(1999)Skadron, Ahuja, Martonosi, and
  Clark]{skadron1999branch}
K.~Skadron, P.~S. Ahuja, M.~Martonosi, and D.~W. Clark.
\newblock Branch prediction, instruction-window size, and cache size:
  {P}erformance trade-offs and simulation techniques.
\newblock \emph{IEEE Transactions on Computers}, 48\penalty0 (11):\penalty0
  1260--1281, 1999.

\bibitem[Smirnov and Tasse(2015)]{tassesmirnov}
O.~Smirnov and C.~Tasse.
\newblock Radio interferometric gain calibration as a complex optimization
  problem.
\newblock \emph{Monthly Notices of the Royal Astronomical Society},
  449\penalty0 (3):\penalty0 2668--2684, 2015.

\bibitem[Strother et~al.(2004)Strother, La~Conte, Hansen, Anderson, Zhang,
  Pulapura, and Rottenberg]{optimize_pipeline}
S.~Strother, S.~La~Conte, L.~K. Hansen, J.~Anderson, J.~Zhang, S.~Pulapura, and
  D.~Rottenberg.
\newblock Optimizing the f{M}{R}{I} data-processing pipeline using prediction
  and reproducibility performance metrics: {I}. {A} preliminary group analysis.
\newblock \emph{Neuroimage}, 23:\penalty0 S196--S207, 2004.

\bibitem[SURF(2018)]{SurfSara}
SURF.
\newblock Grid at {S}{U}{R}{F}sara.
\newblock \url{https://www.surf.nl/en/services-and-products/grid/index.html },
  2018.

\bibitem[Tingay et~al.(2013)Tingay, Goeke, Bowman, Emrich, Ord, Mitchell,
  Morales, Booler, Crosse, Wayth, et~al.]{mwa2}
S.~J. Tingay, R.~Goeke, J.~D. Bowman, D.~Emrich, S.~Ord, D.~A. Mitchell, M.~F.
  Morales, T.~Booler, B.~Crosse, R.~Wayth, et~al.
\newblock The {M}urchison {w}idefield {a}rray: {T}he {s}quare {k}ilometre
  {a}rray {p}recursor at low radio frequencies.
\newblock \emph{Publications of the Astronomical Society of Australia}, 30,
  2013.

\bibitem[Van~Haarlem et~al.(2013)Van~Haarlem, Wise, Gunst, Heald, McKean,
  Hessels, De~Bruyn, Nijboer, Swinbank, Fallows, et~al.]{LOFAR}
M.~Van~Haarlem, M.~Wise, A.~Gunst, G.~Heald, J.~McKean, J.~Hessels,
  A.~De~Bruyn, R.~Nijboer, J.~Swinbank, R.~Fallows, et~al.
\newblock {L}{O}{F}{A}{R}: The low-frequency array.
\newblock \emph{Astronomy \& astrophysics}, 556:\penalty0 A2, 2013.

\bibitem[Van~Weeren et~al.(2016)Van~Weeren, Williams, Hardcastle, Shimwell,
  Rafferty, Sabater, Heald, Sridhar, Dijkema, Brunetti,
  et~al.]{lofar_prefactor}
R.~Van~Weeren, W.~Williams, M.~Hardcastle, T.~Shimwell, D.~Rafferty,
  J.~Sabater, G.~Heald, S.~Sridhar, T.~Dijkema, G.~Brunetti, et~al.
\newblock L{O}{F}{A}{R} facet calibration.
\newblock \emph{The Astrophysical Journal Supplement Series}, 223\penalty0
  (1):\penalty0 2, 2016.

\bibitem[V{\"o}ckler et~al.(2006)V{\"o}ckler, Mehta, Zhao, Deelman, and
  Wilde]{kickstart}
J.-S. V{\"o}ckler, G.~Mehta, Y.~Zhao, E.~Deelman, and M.~Wilde.
\newblock Kickstarting remote applications.
\newblock In \emph{2nd International Workshop on Grid Computing Environments},
  pages 1--8, 2006.

\bibitem[Williams et~al.(2016)Williams, Van~Weeren, R{\"o}ttgering, Best,
  Dijkema, de~Gasperin, Hardcastle, Heald, Prandoni, Sabater,
  et~al.]{Wendy_bootes}
W.~Williams, R.~Van~Weeren, H.~R{\"o}ttgering, P.~Best, T.~Dijkema,
  F.~de~Gasperin, M.~Hardcastle, G.~Heald, I.~Prandoni, J.~Sabater, et~al.
\newblock L{O}{F}{A}{R} 150-{M}{H}z observations of the {B}o{\"o}tes field:
  catalogue and source counts.
\newblock \emph{Monthly Notices of the Royal Astronomical Society},
  460\penalty0 (3):\penalty0 2385--2412, 2016.

\bibitem[Wu et~al.(2013)Wu, Wicenec, Pallot, and Checcucci]{mwa_data_size}
C.~Wu, A.~Wicenec, D.~Pallot, and A.~Checcucci.
\newblock Optimising {N}{G}{A}{S} for the {M}{W}{A} {A}rchive.
\newblock \emph{Experimental Astronomy}, 36\penalty0 (3):\penalty0 679--694,
  2013.

\end{thebibliography}

\end{document}